\documentclass[iop,apj,twocolumn,tighten,twocolappendix,numberedappendix]{emulateapj}
\usepackage{latexsym,graphicx,rotating,amsmath, epsfig, natbib}
\renewcommand{\vec}{\boldsymbol}

\newcommand{\del}{\nabla}

\newcommand{\scrD}{\mathcal{D}}

\newcommand{\scrO}{\mathcal{O}}
\newcommand{\scrR}{\mathcal{R}}

\newcommand{\ms}{\mathrm{m}\thinspace \mathrm{s}^{-1}}

\renewcommand{\dot}{\vec{\cdot}}
\newcommand{\grad}{\vec{\del}}
\renewcommand{\div}{ \grad \dot }

\newcommand{\dV}{\,\mathrm{d}^{3}\vec{x}}
\newcommand{\dt}{\,\mathrm{d}t }

\newcommand{\pd}[2]{\frac{\partial {#1} }{\partial {#2} }}

\newcommand{\Beq}{\begin{eqnarray}}
\newcommand{\Eeq}{\end{eqnarray}}

\newcommand{\eq}[1]{equation~(\ref{#1})}
\newcommand{\eqs}[2]{equations~(\ref{#1})~\&~(\ref{#2})}

\newcommand{\Eq}[1]{Equation~(\ref{#1})}
\newcommand{\Eqs}[2]{Equations~(\ref{#1})~\&~(\ref{#2})}

\newcommand{\pomega}{\varpi}

\usepackage[usenames, dvipsnames]{color}

\shorttitle{Anelastic Gravity Waves}
\shortauthors{Brown, Vasil \& Zweibel} 
\slugcomment{accepted for publication in ApJ}

\begin{document}

  \slugcomment{Accepted for publication in ApJ}

  \received{Jan 26, 2012}
  \accepted{Jul 10, 2012}

  \title{Energy Conservation and Gravity Waves in Sound-proof
    Treatments of Stellar Interiors: Part I Anelastic Approximations}

  \author{Benjamin P.\ Brown}
  \affil{Dept.\ Astronomy, University of Wisconsin, Madison, WI 53706-1582}
  \affil{Center for Magnetic Self Organization in Laboratory and Astrophysical Plasmas, University of Wisconsin, 1150 University Avenue, Madison, WI 53706, USA}
  \email{bpbrown@astro.wisc.edu}
 \author{Geoffrey M.\ Vasil}
  \affil{Canadian Institute for Theoretical Astrophysics, University of Toronto, 60 St.~George Street, Toronto, ON M5S~3H8 Canada}
  \author{Ellen G.\ Zweibel}
  \affil{Dept.\ Astronomy, University of Wisconsin, Madison, WI 53706-1582}
  \affil{Center for Magnetic Self Organization in Laboratory and Astrophysical Plasmas, University of Wisconsin, 1150 University Avenue, Madison, WI 53706, USA}

  \begin{abstract}
    Typical flows in stellar interiors are much slower than the speed
    of sound.  To follow the slow evolution of subsonic motions,
    various sound-proof equations are in wide use, particularly in stellar
    astrophysical fluid dynamics.  These low-Mach number equations
    include the anelastic equations.  Generally, these equations are valid in nearly
    adiabatically stratified regions like stellar convection zones,
    but may not be valid in the sub-adiabatic, stably stratified
    stellar radiative interiors.  Understanding the coupling between
    the convection zone and the radiative interior is a problem of
    crucial interest and may have strong implications for solar and
    stellar dynamo theories as the interface between the two, called
    the tachocline in the Sun, plays a crucial role in many solar
    dynamo theories.  Here we study the properties of gravity waves in
    stably-stratified atmospheres.  In particular, we explore how
    gravity waves are handled in various sound-proof equations.
    We find that some anelastic treatments fail to conserve energy in
    stably-stratified atmospheres, instead conserving pseudo-energies
    that depend on the stratification, and we demonstrate this
    numerically.  One anelastic equation set does conserve energy in
    all atmospheres and we provide recommendations 
    for converting low-Mach number anelastic codes to this set of equations. 

 \end{abstract}
\slugcomment{}


\section{Introduction \& motivation}\label{intro}
In astrophysical fluid dynamics, the evolution time of the fluid flow
is often substantially longer than the sound crossing time of the
system.  This is particularly true for convection deep in stellar
interiors where the flows are very subsonic.  Near the base of the
solar convection zone the sound speed is about 220 km/s, while the
convective velocities are likely of order hundreds of meters per
second.  Following the evolution of sound directly imposes crippling
computational limits on simulations of such flows, as their evolution
times are typically many convective turnover times, each of which is
often several thousand sound times.

So called ``sound-proof'' equations address this separation of scales
by beginning with the Navier-Stokes equations and filtering out fast,
high-frequency sound waves while retaining compressible motions on 
slower time scales due to gravitational stratification.  These motions
include gravity waves in stably stratified regions and asymmetric
convection in unstably stratified regions, with typically broad slow
upflows and narrow fast downflows.  In astrophysical and geophysical
settings,  the most commonly employed ``sound-proof'' equations are
the anelastic equations 
\citep{Batchelor_1953, Ogura&Phillips_1962, Gough_1969}.
These have been employed in various astrophysical and geophysical
codes to study solar convection and the solar dynamo 
\citep[e.g.,][]{Gilman&Glatzmaier_1981, Glatzmaier_1984,
  Glatzmaier_1985, Clune_et_al_1999, 
  Miesch_et_al_2000, Elliott_et_al_2000, Brun&Toomre_2002, Brun_et_al_2004},
stellar convection and dynamos
\citep[e.g.,][]{Browning_et_al_2004, Brun_et_al_2005,
  Brown_et_al_2008, Brown_et_al_2010, Brown_et_al_2011, Nelson_et_al_2011},
the buoyant rise of magnetic structures
\citep[e.g.,][]{Lantz&Fan_1999},
terrestrial convection and the geodynamo
\citep[e.g.,][]{Braginsky&Roberts_1995, Glatzmaier&Roberts_1996,
  Roberts&Glatzmaier_2000, 
  Olson&Christensen_2006, Jones_et_al_2009_JFM, Jones_et_al_2009b}
and the coupling of an unstably stratified convection zone to a stably
stratified region beneath
\citep[e.g.,][]{Rogers_et_al_2003, Rogers&Glatzmaier_2005_ApJ,
  Rogers&Glatzmaier_2005_MNRAS, Rogers&Glatzmaier_2006, 
  Browning_et_al_2006,  
  Rogers_et_al_2006, Rogers_et_al_2008, 
  Rogers&MacGregor_2011, Brun_et_al_2011}.  
Recently a significant benchmarking effort has been undertaken to
compare the various implementations of the anelastic equations \citep{Jones_et_al_2011}.

Formally the anelastic approximation is only valid for an adiabatic or
nearly adiabatic 
atmosphere.  The solar convection zone is nearly adiabatic but it is
underlain by a stably stratified radiative zone; unsurprisingly the
anelastic equations are often extended into this region where their
validity may break down, to study the coupling of penetrative
convection with a stably-stratified region
\citep[e.g.,][]{Rogers&Glatzmaier_2006, Rogers_et_al_2008,  Rogers&MacGregor_2010, Rogers&MacGregor_2011, Brun_et_al_2011}.
This is particularly important in simulations of the solar dynamo, as
the stably stratified internal boundary layer known as the tachocline
at the base of the convection zone is thought to play a major role in
the global-scale dynamo.

Fundamentally, the anelastic equations filter sound waves by modifying
the continuity equation of the fully compressible Navier-Stokes
equations.  Questions about the energy conserving properties of the
anelastic approximation have remained a thorny issue in the fluid
dynamics community, with an especially vigorous debate occurring in the
atmospheric sciences \citep[e.g.,][]{Durran_1989, Bannon_1996}, 
where these equations were originally derived.  Likewise, there are
several competing anelastic approaches, including 
``co-density'' formulations \citep[e.g.,][ hereafter the LBR equations]{Lantz_1992, Braginsky&Roberts_1995} 
and their different properties are unclear.  

An alternate approach to sound-proofing the Navier-Stokes
equations are the pseudo-incompressible equations, where the pressure
rather than continuity equation is modified.  
These equations were proposed in \citep{Durran_1989} and
have recently been adopted in the astrophysical fluid dynamics
community \citep[e.g.,][]{Almgren_et_al_2006a_ApJ, 
Almgren_et_al_2006b_ApJ, Zingale_et_al_2009} and see particular use
in the MAESTRO code \citep{Nonaka_et_al_2010}.  
The properties of gravity waves and stable-layer dynamics in the
pseudo-incompressible equations have been explored extensively in the
atmospheric sciences community, with several comparisons against the
properties of the anelastic equations
\citep{Durran_1989, Durran_2008, Nance&Durran_1994, Achatz_et_al_2010,
Klein_et_al_2010}.  We reserve further discussion of gravity waves in
this set of equations for a later paper.

Here we explore three implementations of the anelastic equations, one
used in the anelastic spherical harmonic (ASH) code, and two different
implementations of the ``co-density'' formulation (LBR equations).  
These equation sets are detailed in Section~\ref{sec:anelastic equations}.  We show that the
anelastic equations based directly on the Navier-Stokes equations (anelastic
Navier-Stokes, or ANS equations) behave incorrectly in stably
stratified region. First we analytically study wave motions in an 
isothermal atmosphere in Section~\ref{sec:isothermal atmosphere}.  
We find that these equations do not conserve energy and instead
conserve an entropy-weighted ``pseudo-energy'' (Section~\ref{sec:self-adjointness}).  
We find however that the LBR equations do behave
correctly for strongly stratified regions, conserving energy and reproducing the results
obtained from the full compressible Euler equations.  This is surprising, as
the LBR equations make further assumptions of adiabaticity beyond
those contained in the basic ANS equations, but these assumptions lie
at the heart of the energy-conserving properties.  As a consequence,
adjustments to the LBR equations to more correctly capture the
sub-adiabatic stratification can have profound consequences,
introducing a completely different form of energy non-conservation 
\citep[e.g., ][and hereafter the RG equations]{Rogers&Glatzmaier_2005_ApJ}.
We explore the behavior of these differing equations further in
bounded atmospheres and spherical geometries in 
Section~\ref{sec:bounded geometries} and perform numerical simulations
that show the difference between the normal ANS equations and the
LBR equations.  The implications of these findings
for simulations of solar convection is discussed in
Section~\ref{sec:conclusions}, which also give suggestions for
improving anelastic treatments of stably-stratified regions.
The reader who is primarily interested in implementing
energy-conserving anelastic equations should read
Sections~\ref{sec:anelastic equations}, \ref{sec:bounded geometries}
and \ref{sec:conclusions}.

\section{Model equations}
\label{sec:anelastic equations}
\subsection{Fully compressible Euler equations}
For the purposes of this paper, the most general equations for fluid
dynamics in the solar interior are the fully compressible
Navier-Stokes equations.  When viscosity is neglected, as we do here,
these are known as the fully compressible Euler equations (FC
equations).  The equations of continuity and momentum are 
\begin{eqnarray}
  \frac{\partial \rho}{\partial t} + \vec{u}\cdot\vec{\del}\rho
  &=& -\rho\vec{\del}\cdot\vec{u} ,
  \label{eq:compressible continuity}
  \label{eq:Euler continuity}
\\
  \rho\left(\frac{\partial \vec{u}}{\partial t} + \vec{u}\cdot\vec{\del}\vec{u}\right)
  &=& -\vec{\del}P  + \rho\vec{g} ,
  \label{eq:compressible momentum}
  \label{eq:Euler momentum}
\end{eqnarray}
with gravitational acceleration $\vec{g} = -g \vec{\hat{r}}$.
For an ideal gas, 
\begin{equation}
P=\scrR \rho T =  (\gamma-1)\rho \mathcal{E} ,
\end{equation}
with $\mathcal{E}$ the specific internal energy and $\gamma=c_P/c_V=5/3$
is the ratio of specific heats.  Here $c_P = \gamma/(\gamma -1) \scrR$
is the specific heat at constant pressure. 
The evolution equations for temperature and pressure are
\begin{eqnarray}
  \frac{\partial T}{\partial t} + \vec{u}\cdot\vec{\del}T &=& - (\gamma - 1 ) T \vec{\del}\cdot\vec{u},
  \label{eq:compressible T}
\\
 \frac{\partial P}{\partial t} + \vec{u}\cdot\vec{\del}P &=& - \gamma P \vec{\del}\cdot \vec{u},
\label{eq:compressible P}
\end{eqnarray}
where thermal conduction and other sources and sinks of energy are neglected.

Although equations~(\ref{eq:compressible continuity}--\ref{eq:compressible P}) 
form a complete system, it will be useful during our discussion of the
anelastic equations to rewrite these in terms of entropy $S$.
Equations~(\ref{eq:compressible T}) and (\ref{eq:compressible P}) can
be combined with an equation of state linking
the thermodynamic properties
\begin{equation}
  \frac{dS}{c_P} = \frac{1}{\gamma} d\ln P - d\ln \rho 
   = \frac{1}{\gamma} d\ln T - \frac{\gamma-1}{\gamma} d\ln \rho
  \label{eq:background EOS}
\end{equation}
into an equation for the evolution of entropy fluctuations
\begin{equation}
  \frac{\partial S}{\partial t} + \vec{u}\cdot\vec{\del}S = 0.
  \label{eq:Euler entropy}
\end{equation}

We now specialize to the case of a hydrostatically balanced,
stratified atmosphere with background density stratification 
$\rho_0$, pressure $P_0$, temperature $T_0$ and entropy $S_0$ that
only vary with radius, with
\begin{equation}
  \vec{\del}P_0  = \rho_0\vec{g}.
  \label{eq:hydrostatic balance}
\end{equation}
We define fluctuating quantities, denoted with
subscript 1, by subtracting the time-independent
hydrostatic atmosphere making no assumptions about relative
amplitudes, with e.g., $P_1 \equiv P - P_0(r)$, thus these equations
are fully nonlinear. 

\newpage
\subsection{Anelastic models and fully compressible Euler equations in standard form}

All anelastic approximations employ a continuity equation of the form
\begin{equation}
  \vec{\del}\cdot\left(\rho_0 \vec{u}\right)=0.
  \label{eq:anelastic continuity}
\end{equation}
Equation~(\ref{eq:anelastic continuity}) derives
from the assumption that the density fluctuations are small
\begin{equation}
  \rho_1 \equiv \rho - \rho_0 \ll \rho_0.
\end{equation}
In this case, the fluctuating density is given by the linearized equation of state, 
\begin{equation}
  \frac{\rho_1}{\rho_0} = \frac{1}{\gamma}\frac{P_1}{P_0} - \frac{S_1}{c_P}  
                                   = \frac{P_1}{P_0} - \frac{T_1}{T_0},
  \label{eq:linearized EOS}
\end{equation}
and though using a linear equation of state is not strictly required, we find it a
clarifying simplification for the current discussion.

We consider equation~(\ref{eq:anelastic continuity}) to be the
defining characteristic of anelastic models.  There exist however a
variety of different treatments for the momentum and energy equations
in the anelastic literature.   In the following subsections we will
consider three common formulations.  The different notation and
different thermodynamics used in the various anelastic treatments
leads to some confusion.  To remedy this, we reproduce each set of
models under as consistent a notation as possible.  Practical
\emph{numerical} or computational differences can arise when
\emph{solving} different transformations of the same fundamental
model, but these issues lie beyond our current scope.  Therefore, we
consider two models identical if one can bring them into the same form
by legitimate mathematical transformation, i.e., without
approximation.

For comparison with the anelastic equations we first write the FC
equations in standard form.  With a linearized equation of
state~(\ref{eq:linearized EOS}), we rewrite the buoyancy term
involving pressure fluctuations in the following fashion
\begin{equation}
\frac{P_1}{\gamma P_0}\vec{g} = \frac{P_1}{\rho_0}
\frac{\vec{\del}P_0}{\gamma P_0} = 
\frac{P_1}{\rho_0} \left[\vec{\del} \left(\frac{S_0}{c_P}\right) + \vec{\del} \ln \rho_0 \right],
\end{equation}
where we have used equations  (\ref{eq:hydrostatic balance}) and (\ref{eq:background EOS}).
We now introduce the reduced or kinematic pressure $\pomega$ with
\begin{equation}
\pomega \equiv \frac{P_1}{\rho_0}.
\end{equation}
The fully compressible Euler equations, with an entropy based energy
equation and reduced pressure $\pomega$, are
\begin{eqnarray}
\frac{\partial \rho_1}{\partial t} + \vec{u}\cdot\vec{\del}\rho_0
  &=& -\rho_0\vec{\del}\cdot\vec{u},
  \label{eq:compressible continuity linearized rho}
\\
 \frac{\partial \vec{u}}{\partial t} + \vec{u}\cdot\vec{\del}\vec{u}
  &=& -\vec{\del} \pomega + \pomega \vec{\del}\left(\frac{S_0}{c_P}\right) - \frac{S_1}{c_P} \vec{g}, 
  \label{eq:compressible momentum pomega}\\
  \frac{\partial S_1}{\partial t} + \vec{u}\cdot\vec{\del}S_1 &=& - \vec{u}\cdot\vec{\del}S_0.
  \label{eq:compressible entropy}
  \label{eq:Euler fluctuating entropy}
\end{eqnarray}
These equations linearize the thermodynamic variables
(eq.~\ref{eq:linearized EOS}) but are nonlinear in the velocities
and are the counterparts of the anelastic equations that we now turn
to; we do not solve these equations 
(\ref{eq:compressible continuity linearized rho}--\ref{eq:Euler fluctuating entropy})
but include them for illustrative purposes.

\subsection{ANS Anelastic equations}

In many anelastic equations the momentum equation is the same as in
the FC equations \citep[e.g.,][]{Gilman&Glatzmaier_1981,
  Drew_et_al_1995, Clune_et_al_1999,  Brun_et_al_2004}. 
We thus refer to these equations as the anelastic Navier-Stokes (ANS) equations.
In the ANS equations, the momentum equation is
\begin{equation}
  \rho_0\left(\frac{\partial \vec{u}}{\partial t} + \left(\vec{u}\cdot\vec{\del}\right)\vec{u}\right)  = -\vec{\del}P_1  +  \rho_1 \vec{g},
 \label{eq: anelastic NS}
\end{equation}
which with equation~(\ref{eq:linearized EOS}) can be transformed into
the same form as equation~(\ref{eq:compressible momentum pomega}), with
\begin{equation}
\frac{\partial \vec{u}}{\partial t} + \vec{u}\cdot\vec{\del}\vec{u}
  = -\vec{\del} \pomega + \pomega
  \vec{\del}\left(\frac{S_0}{c_P}\right) - \frac{S_1}{c_P} \vec{g} .
  \label{eq:ASH momentum pomega}
\end{equation}
The ANS momentum equation~(\ref{eq:ASH momentum pomega}) can be
written in an alternative form by combining $\pomega$ terms to yield
\begin{equation}
\frac{\partial \vec{u}}{\partial t} + \vec{u}\cdot\vec{\del}\vec{u}
  = - e^{(S_0/c_P)} \vec{\del}\left(\pomega e^{-(S_0/c_P)}\right) - \frac{S_1}{c_P} \vec{g},
  \label{eq:ASH momentum pomega energy form}
\end{equation}
which will be useful for our analysis in Section~\ref{sec:self-adjointness}.  
As a notational issue, the stratification term interacting with
$\pomega$ in equation~(\ref{eq:ASH momentum pomega energy form}) 
takes the same form as a potential  temperature, as is
traditionally used in studies of geophysical flows in the atmosphere
and ocean with 
\begin{equation}
  \Theta_0 \equiv e^{(S_0/c_P)} = \frac{P_0^{1/\gamma}}{\rho_0}
\end{equation}
and, with the linearized equation of state~(\ref{eq:linearized EOS}),
\begin{equation}
  \frac{\Theta_1}{\Theta_0} = \frac{S_1}{c_P}.
\end{equation}
In terms of $\Theta$, the ANS momentum equation is
\begin{equation}
\frac{\partial \vec{u}}{\partial t} + \vec{u}\cdot\vec{\del}\vec{u}
  = - \Theta_0 \vec{\del}\left(\pomega \Theta_0^{-1}\right)
      -\frac{\Theta_1}{\Theta_0} \vec{g}.
  \label{eq:ASH momentum pomega energy form potential T}
\end{equation}

Neglecting diffusion and sources of energy, the energy equation is the
same as in the FC equations (eq.~\ref{eq:Euler entropy}), with a
background entropy gradient
\begin{equation}
  \frac{\partial S_1}{\partial t} + \vec{u}\cdot\vec{\del}S_1 = - \vec{u}\cdot\vec{\del}S_0.
  \label{eq:ASH entropy}
\end{equation}
Combined with the anelastic continuity equation~(\ref{eq:anelastic continuity}), 
equations~(\ref{eq:ASH momentum pomega}) and (\ref{eq:ASH entropy})
constitute a full set of equations for anelastic motions.

\subsection{LBR Anelastic equations}
In the ``co-density'' equations or Lantz-Braginsky-Roberts
equations \citep[e.g.,][and hereafter LBR equations]{Lantz_1992, Braginsky&Roberts_1995,
  Lantz&Fan_1999, Jones_et_al_2009_JFM}, the $\pomega
  \vec{\del}(S_0/c_P)$ term is dropped and the momentum equation becomes
\begin{equation}
  \frac{\partial \vec{u}}{\partial t} + \vec{u}\cdot\vec{\del}\vec{u}
  = -\vec{\del}\pomega  - \frac{S_1}{c_P} \vec{g} ,
 \label{eq:LBR momentum pomega}
\end{equation}
or, in terms of potential temperatures, 
\begin{equation}
\frac{\partial \vec{u}}{\partial t} + \vec{u}\cdot\vec{\del}\vec{u}
  = \vec{\del}\pomega
      -\frac{\Theta_1}{\Theta_0} \vec{g}.
  \label{eq:LBR momentum pomega energy form potential T}
\end{equation}
As in the ANS equations, the energy equation~(\ref{eq:ASH entropy})
and the continuity equation~(\ref{eq:anelastic continuity}) complete
the full set of equations.  

The LBR momentum equation~(\ref{eq:LBR momentum pomega}) 
is derived from the full Euler momentum equation~(\ref{eq:Euler momentum}) 
by assuming $\vec{\del}  S_0 \approx 0$, as for
a nearly adiabatic state.  Despite this assumption, we find that the
LBR equations perform well when $\vec{\del} S_0  \neq 0$
while the ANS equations perform poorly in that
same limit.   Though \cite{Lantz_1992} and \cite{Braginsky&Roberts_1995} 
are typically credited with independently deriving the LBR
equations, these equations bear striking similarities to the
Lipps-Hemler anelastic equations \citep{Lipps&Hemler_1982,  
Lipps&Hemler_1985, Lipps_1990}, who were possibly the first to
recognize the importance of introducing a reduced pressure and
neglecting the interactions between fluctuating pressure and
stratification.  They likewise recognized that gravity waves derived
from their anelastic equations conserved energy.

\subsection{RG anelastic equations}
\cite{Rogers&Glatzmaier_2005_ApJ} use a different set of
anelastic equations (hereafter the RG equations).  As above, 
neglecting viscosity and sources of heat, their equations are
the momentum equation and a temperature based energy equation 
\begin{eqnarray}
  \frac{\partial \vec{u}}{\partial t} + \vec{u}\cdot\vec{\del}\vec{u}
 &=& -\vec{\del}\pomega + \pomega \vec{\del} \ln T_0 - \frac{T_1}{T_0} \vec{g}, 
 \label{eq:RG momentum pomega}
\\
\frac{\partial T_1}{\partial t} + \vec{u}\cdot\vec{\del}T_1 &=& 
-\vec{u}\cdot\vec{\del}T_0 - (\gamma - 1 )(T_0+T_1)\vec{\del}\cdot\vec{u} \ .\nonumber\\
\end{eqnarray}
With the anelastic continuity equation~(\ref{eq:anelastic continuity}), 
these constitute a full set of equations for anelastic motions.

Equation~(\ref{eq:RG momentum pomega}) can be equivalently written
\begin{equation}
 \frac{\partial \vec{u}}{\partial t} + \vec{u}\cdot\vec{\del}\vec{u}
   = -T_0\vec{\del}\left(\pomega/T_0\right) - \frac{T_1}{T_0}\vec{g},
  \label{eq:RG momentum pomega energy form}
\end{equation}
a form that will be useful in Section~\ref{sec:self-adjointness}.
By combining the equation of state~(\ref{eq:background EOS}) with the
anelastic continuity equation~(\ref{eq:anelastic continuity}), we can
cast the energy equation in terms of entropy with
\begin{equation}
  \left(\frac{\partial}{\partial t} + \vec{u}\cdot\vec{\del}\right)\frac{T_1}{T_0} = 
  -\gamma \left(1 + \frac{T_1}{T_0}\right) \vec{u}\cdot\vec{\del}\left(S_0/c_P\right).
  \label{eq:RG energy entropy}
\end{equation}
With a linearized equation of state, this takes the same form as the
entropy equation~(\ref{eq:compressible entropy}), but with an extra
factor of $\gamma$ multiplying the background entropy gradient.

The right hand sides of the momentum equations for these four systems
of equations are summarized in Table~\ref{table:RHS momentum}.

\begin{deluxetable}{clccccccccccc}
 \tabletypesize{\footnotesize}
  \tablecolumns{3}
  \tablewidth{0pt}  
  \tablecaption{Systems of equations
  \label{table:RHS momentum}}
  \tablehead{\colhead{System}  &  
    \colhead{RHS momentum equation} &
    \colhead{eq} 
}
 \startdata
 FC &
$ -\vec{\del} \pomega 
  + \pomega\vec{\del}\left(S_0/c_P\right) - \left(S_1/c_P\right) \vec{g}$
& (\ref{eq:compressible momentum pomega}) \\[2mm]
 ANS &
$ -\vec{\del} \pomega 
  + \pomega\vec{\del}\left(S_0/c_P\right) - \left(S_1/c_P\right) \vec{g}$
& (\ref{eq:ASH momentum pomega}) \\[2mm]
 LBR & 
$ -\vec{\del} \pomega - \left(S_1/c_P\right)\vec{g}$
& (\ref{eq:LBR momentum pomega}) \\[2mm]
 RG & 
$-\vec{\del}\pomega + \pomega \vec{\del}\ln T_0 - (T_1/T_0) \vec{g}$
& (\ref{eq:RG momentum pomega}) \\
\cutinhead{RHS wave momentum}
FC &
$ -\vec{\del} \pomega 
  + \pomega\vec{\del}\left(S_0/c_P\right) +  (\vec{\xi}\cdot \vec{\del}) (S_0/c_P)\vec{g}$
& 
\\[2mm]
 ANS &
$ -\vec{\del} \pomega 
  + \pomega\vec{\del}\left(S_0/c_P\right)  + (\vec{\xi}\cdot \vec{\del}) (S_0/c_P) \vec{g} $
& 
\\[2mm]
 LBR & 
$ -\vec{\del} \pomega + (\vec{\xi}\cdot \vec{\del}) (S_0/c_p) \vec{g} $
& 
\\[2mm]
 RG & 
$-\vec{\del}\pomega + \pomega \vec{\del}\ln T_0 +\gamma(\vec{\xi}\cdot\vec{\del}) (S_0/c_P) \vec{g}$
& 
\\
\cutinhead{RHS momentum (for Section~\ref{sec:self-adjointness})}
 ANS &
$ -e^{S_0/c_P}\vec{\del} \left(\pomega e^{-S_0/c_P}\right)
   -\left(S_1/c_P\right) \vec{g}$
& (\ref{eq:ASH momentum pomega}) \\[2mm]
 LBR & 
$ -\vec{\del} \pomega - \left(S_1/c_P\right)\vec{g}$
& (\ref{eq:LBR momentum pomega}) \\[2mm]
 RG & 
$-T_0\vec{\del}(\pomega/T_0) - (T_1/T_0) \vec{g}$
& (\ref{eq:RG momentum pomega}) 
 \enddata
\tablecomments{In all systems of equations, $\pomega=P_1/\rho_0$.  
  The fully compressible equations use continuity
  equation~(\ref{eq:compressible continuity}) while anelastic systems
  use equation~(\ref{eq:anelastic continuity}).  In the wave momentum equations,
  $\vec{\xi}$ is the displacement vector as defined in eq~\ref{eq:xi}.
}
\end{deluxetable}

\section{Linear Waves in an isothermal atmosphere}
\label{sec:isothermal atmosphere}

A plane-parallel isothermal atmosphere gives an analytically tractable
background for computing eigenfrequencies and modes for linear
gravity and/or acoustic waves.  Computing these simple solutions 
helps elucidate the differences between the various anelastic treatments.   
Defining the velocity in terms of the vector displacement,
\begin{equation}
  \vec{u} = \frac{\partial \vec{\xi}}{\partial t} 
  \label{eq:xi}
\end{equation}
allows simple integration of the linear thermodynamic equations,
\begin{eqnarray}
  \rho_1/\rho_0 &=& -\vec{\xi}\cdot\vec{\del} \ln \rho_0 -  \vec{\del}\cdot\vec{\xi}, 
  \label{eq:frozen-in density} \\
  P_1/\gamma P_0 &=& -\vec{\xi}\cdot\vec{\del} \ln P_0^{1/\gamma} -  \vec{\del}\cdot\vec{\xi}, 
  \label{eq:frozen-in pressure}\\
  S_1 &=& -\vec{\xi}\cdot\vec{\del} S_0.
  \label{eq:frozen-in entropy}
\end{eqnarray}
For wavelike perturbations in an atmosphere of infinite
extent, we can assume without loss of generality that
\begin{equation}
  (\vec{\xi}, \rho_1, P_1, S_1) \propto f(K r) \exp{\left(i\omega t  - i m x\right)} 
\label{eq:wave properties}
\end{equation}
where $x$ is the horizontal coordinate, $m$ is the horizontal wave
number, and the vertical dependence on $r$ has been left in general
form with wavenumber $K$.

In a hydrostatically balanced isothermal atmosphere
\begin{equation}
  \del_r \ln P_0 = \del_r \ln \rho_0 = -\frac{g}{\scrR T_0} =
  -\frac{1}{H} =  -\frac{\gamma g}{c_S^2}
  \label{eq:isothermal scaleheight}
\end{equation}
where $H$ is the pressure or density scale height, $\del_r$ is the
vertical derivative, and
\begin{equation}
  c_S^2 \equiv \frac{\gamma P_0}{\rho_0} 
\end{equation}
is the (constant) sound speed.
The Brunt-V\"ais\"al\"a frequency $N$ is
\begin{equation}
  N^2 \equiv -\vec{g} \cdot \vec{\del} \left(\frac{S_0}{c_P} \right)=
  \frac{(\gamma -1)}{\gamma} \frac{g}{H}.
\label{eq:N2}
\end{equation}

\subsection{Fully compressible equations}  
 
The solution for the fully compressible equations is well known and
can be found in several textbooks \citep[e.g.,][]{Lighthill_1978_Waves_in_Fluids}.
We begin with the linearized momentum equation for waves
\begin{equation}
  -\rho_0\omega^2 \vec{\xi} = -\vec{\del}P_1 + \rho_1 \vec{g},
  \label{eq:linear Euler}
\end{equation}
and solve for eigenfrequencies using 
equations~(\ref{eq:xi})--(\ref{eq:frozen-in pressure}) and 
(\ref{eq:wave properties}).
Taking the vertical eigenfunction $f(K r) = \exp{( - i K r)}$, the
dispersion relationship for waves in an isothermally-stratified atmosphere is 
\begin{equation}
  -\frac{\omega^4}{c_S^2} +\omega^2 \left[K^2 + m^2 - i K H^{-1}  \right] = m^2 N^2.
\label{eq:compressible dispersion I}
\end{equation}
It is well known that the fully compressible Euler equations conserve
energy.  Their frequencies $\omega$ must thus be
purely real with no imaginary component (see Section~\ref{sec:self-adjointness}), yet
equation~(\ref{eq:compressible dispersion I}) has an imaginary
component.  Taking a complex vertical wavenumber
\begin{equation}
K = k + i\frac{1}{2 H}
\label{eq:compressible vertical wavenumber}
\end{equation}
with real component $k$ resolves this.  The vertical eigenfunction
follows
\begin{equation}
  f(K r) = \exp{\left(- i K r\right)} =
  \exp{\left(\frac{r}{2H}\right)}\exp{\left( - i k r\right)}. 
\label{eq:compressible vertical eigenfunction}
\end{equation}
All waves in this atmosphere share the properties that their
eigenfunctions grow with height, their momentum density decreases with height
\begin{equation}
  \rho_0 u \propto \exp{\left(-\frac{r}{2H}\right)},
\end{equation}
while their kinetic energy $\rho_0 u^2$ is constant with height.
These eigenfunctions are orthogonal with respect to the density weight
\begin{equation}
  \int \rho_0 f(K r) f(K^\prime r)^* dr = \delta(k-k^\prime),
\end{equation}
with $\delta$ here the Dirac delta. 

The final dispersion relationship with $\omega^2$ real is
\begin{equation}
  -\frac{\omega^4}{c_S^2} +\omega^2 \left[k^2 + m^2 +\frac{1}{4H^2}\right] = m^2 N^2.
\label{eq:compressible dispersion}
\end{equation}
with $N^2$ given by equation~(\ref{eq:N2}).
The quadratic nature of equation~(\ref{eq:compressible dispersion})
in $\omega^2$ provides for two distinct branches of acoustic and
gravity waves.  In the high frequency limit $\omega^2 \gg N^2$, 
\begin{equation}
  \omega_{SW}^2 = \left[k^2 + m^2 +\frac{1}{4H^2}\right]  c_S^2,
\end{equation}
representing the propagation of pure sound waves in an atmosphere
with an acoustic cutoff frequency $c_S^2/4H^2$.  In the low frequency
limit, we obtain pure internal gravity waves with
\begin{equation}
  \omega_{GW}^2 = \frac{m^2}{k^2 + m^2 +\frac{1}{4H^2}} N^2.
  \label{eq:compressible dispersion GW}
\end{equation}
The full solution for $\omega^2$ follows
\begin{equation}
  \omega^2 = \frac{\omega^2_{SW}}{2} \left( 1\pm \sqrt{1-\frac{4 \omega^2_{GW}}{\omega^2_{SW}}}\right)
  \label{eq:omega Navier-Stokes infinite isothermal full dispersion}
\end{equation}
with the positive root corresponding to the sound waves while the negative
root corresponds to the internal gravity waves.

\subsection{ANS gravity waves}
\label{sec:ANS waves}
We begin our analysis of the anelastic systems with the ANS equations.
For linear waves, the continuity, momentum and
energy equations are
\begin{eqnarray}
\vec{\del}\cdot\vec{\xi} &=& -\vec{\xi} \cdot \vec{\del}\ln \rho_0 ,
\label{eq:anelastic wave continuity}
\\
-\omega^2 \vec{\xi} &=& -\vec{\del} \pomega +
\pomega \vec{\del}\left(\frac{S_0}{c_P}\right) -
\frac{S_1}{c_P}\vec{g},
\label{eq:anelastic wave momentum}
 \\
S_1 &=& -\vec{\xi} \cdot \vec{\del} S_0. 
\label{eq:anelastic wave energy}
\end{eqnarray}
Combining the vertical momentum equation~(\ref{eq:anelastic wave
  momentum}) and the energy equation~(\ref{eq:anelastic wave energy})
for linearized waves, we obtain
\begin{equation}
 \omega^2 \xi_r -\del_r (\pomega) + \pomega \del_r (S_0/c_P) =
    N^2 \xi_r,
   \label{eq:ASH vertical momentum omega}
\end{equation}
where we have also used equation~(\ref{eq:N2}).
We obtain $\pomega$ by taking the horizontal divergence of the
momentum equation
\begin{equation}
\omega^2\vec{\del_\perp}\cdot \vec{\xi} = \del^2_\perp \pomega.
\label{eq:pomega horizontal divergence}
\end{equation}

The dispersion relationship for linear waves is
\begin{equation}
\omega^2\Big[m^2+ \left(K - i \frac{2\gamma -1}{2 \gamma H}\right)^2 +
\frac{1}{4 \gamma^2 H^2} \Big] =  m^2 N^2.
\end{equation}
Once again, $\omega^2$ has an imaginary component.  As with the fully
compressible equations, we can try to absorb this imaginary component
within the vertical eigenfunction, which leads to a vertical wave number
\begin{equation}
  K = k + i \frac{1}{2H} \frac{2 \gamma -1}{\gamma},
  \label{eq:ASH vertical wavenumber real omega}
\end{equation}
and vertical eigenfunction
\begin{equation}
  f(K r) = \exp{\left(-i K r\right)} = \exp{\left( \frac{r}{2H}\frac{2\gamma -1}{\gamma}\right)}
\exp{\left(- i k r\right)},
  \label{eq:ASH vertical eigenfunction real omega}
\end{equation}
with time dependence
\begin{equation}
   \omega_\mathrm{ANS}^2\Big[m^2+ k^2 +\frac{1}{4\gamma^2 H^2}\Big] = m^2 N^2.
   \label{eq:ASH dispersion relationship real omega}
\end{equation}
A serious problem lurks within these choices however, as the
momentum and kinetic energy densities scale as 
\begin{eqnarray}
  \rho_0 u &\propto& \exp\left(-\frac{1}{2 \gamma}\frac{r}{H}\right),\\
  \rho_0 u^2 &\propto& \exp\left(\frac{\gamma -1}{\gamma} \frac{r}{H}\right).
\end{eqnarray}
For adiabatic motions in an ideal gas, $\gamma = 5/3$, and the
kinetic energy of the waves grows exponentially with height.

Alternatively, we can use the eigenfunctions from the
fully compressible equations, in 
equations~(\ref{eq:compressible vertical eigenfunction}-\ref{eq:compressible vertical wavenumber}), 
which leads to the
correct far-field behavior for momentum and energy, but leads to a
dispersion relationship of
\begin{equation}
   \omega_\mathrm{ANS-I}^2\Big[m^2+ k^2 +\frac{1}{4 H^2}\frac{2 - \gamma}{\gamma} 
   + i k g \left(\gamma - 1\right)\Big] = m^2 N^2.
   \label{eq:ASH dispersion relationship}
\end{equation}
There is now an imaginary component to $\omega^2$ and anelastic
gravity waves in an infinite isothermal atmosphere can have spurious
growing (or decaying) modes.  As we will see in
Section~\ref{sec:self-adjointness}, this bizarre behavior reflects the
fact that the ANS equations do not conserve energy.  Further, as we
will see in Sections~\ref{sec:self-adjointness}--\ref{sec:bounded geometries},
 the fact that these spurious modes have not been detected in simulations previously
is likely related to the presence of a conserved pseudo-energy (i.e.,
a differently weighted quadratic integral of the fluctuating velocities and entropies).

\begin{figure}
  \begin{center}
    \includegraphics[width=\linewidth]{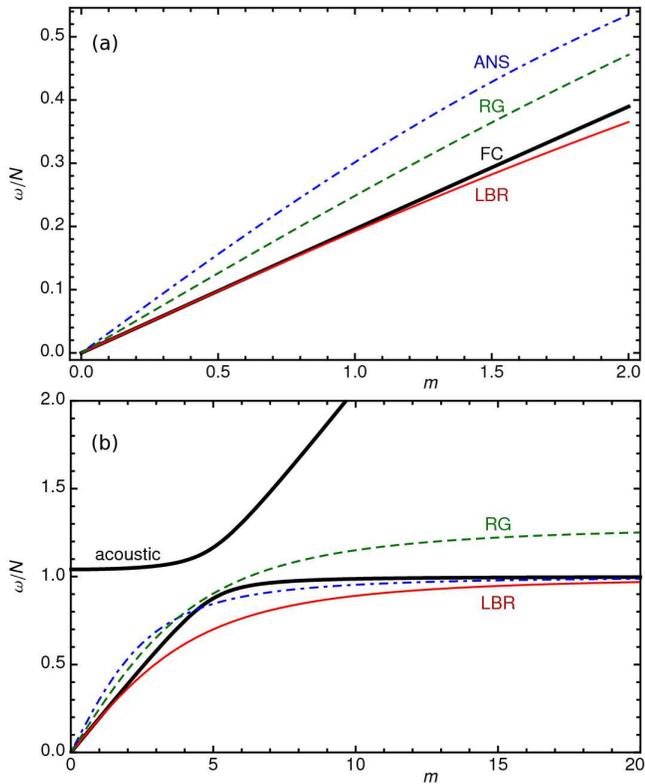}
  \end{center}
\caption{Dispersion relationships for gravity waves in an isothermal
  atmosphere of infinite extent with $1/H=10$, for the fundamental
  mode $k=1$ and  with horizontal wavenumber $m$.  
  $(a)$ Frequencies for each set of equations in the low frequency
  limit ($\omega^2/N^2 \ll 1)$.
  Shown are the gravity wave branch of the exact solutions for the
  fully compressible Euler equations (black, labelled FC, and given by
  eq~\ref{eq:omega Navier-Stokes infinite isothermal full dispersion}).  
  Also shown are the dispersion relation for the ANS equations 
  (blue, dot-dashed), the LBR equations (red, solid),
  and the RG equations (green, dashed) with each line labeled.
  In this regime the LBR equations and exact solutions to the Euler
  (FC) equations are in good agreement, 
  while the ANS and RG equations obtain frequencies that are too high.
  The corresponding dispersion relationships are given in 
  Table~\ref{table:dispersion relationships}.
  $(b)$ Full frequency domain.  At large $m$ the ANS, LBR and FC
  equations converge to the Brunt-V\"ais\"al\"a frequency $N$, while
  the RG equations are too large by a factor of $\sqrt{\gamma}$.  
  Here we also show the sound wave branch
  (black, labelled ``acoustic'', and given by 
  eq~\ref{eq:omega Navier-Stokes infinite isothermal full dispersion}) 
  of the exact solution to the full Euler equations. 
  \label{fig:omega infinite atm}
}
\end{figure}

\subsection{LBR gravity waves}
Finding linear eigenfrequencies in the LBR equations amounts to the
same procedure as in Section~\ref{sec:ANS waves}.  
Now however the $\pomega \vec{\del}(S_0/c_P)$ term is missing from the
vertical momentum equation, and equation~(\ref{eq:ASH vertical momentum omega})
becomes
\begin{equation}
\omega^2 \xi_r -\del_r \pomega = N^2 \xi_r.
\label{eq:LBR vertical momentum omega}
\end{equation}
This readily yields the following dispersion relationship
\begin{equation}
\omega^2 \left[ m^2 + \left(K-i\frac{1}{2H}\right)^2 + \frac{1}{4 H^2} \right] = m^2 N^2,
\label{eq:LBR vertical momentum omega isothermal final}
\end{equation}
Requiring that $K=k+i/2H$ is clearly the natural choice for obtaining
real frequencies.  By employing the vertical eigenfunctions in
equations~(\ref{eq:compressible vertical eigenfunction}-\ref{eq:compressible vertical wavenumber})
we get
\begin{equation}
  \omega_\mathrm{LBR}^2 \left[ m^2 + k^2 + \frac{1}{4 H^2}\right] = m^2 N^2,
  \label{eq:LBR dispersion relationship}
\end{equation}
which is the same as equation~(\ref{eq:compressible dispersion GW}).
Adiabatically propagating gravity waves solved with the LBR equations
in an infinite isothermal atmosphere behave like the low frequency
branch of the fully compressible equations in both their time
dependence and their vertical structure.

\subsection{RG gravity waves}
Next we look at the propagation of gravity waves within the
RG equations.  In an isothermal atmosphere the coupling
between $\pomega$ and the background stratification disappears.  With
the anelastic continuity equation~(\ref{eq:anelastic wave continuity}), the
linear RG wave equations are
\begin{eqnarray}
-\omega^2 \vec{\xi} &=& -\vec{\del} \pomega - \frac{T_1}{T_0}\vec{g},
\label{eq:RG wave momentum}
 \\
\frac{T_1}{T_0} &=& -\gamma \vec{\xi} \cdot \vec{\del} (S_0/c_P).
\label{eq:RG wave energy}
\end{eqnarray}
Combining the vertical momentum and energy equations yields
\begin{equation}
\omega^2 \xi_r  -\del_r \pomega = \gamma N^2 \xi_r
 \label{eq:RG vertical momentum omega}
\end{equation}
This leads to a dispersion relationship of
\begin{equation}
\omega^2\left[ m^2 + \left(K - i\frac{1}{2H}\right)^2 + \frac{1}{4H^2}\right] = \gamma m^2 N^2.
\end{equation}
As previously, the vertical eigenfunctions in
equations~(\ref{eq:compressible vertical eigenfunction}-\ref{eq:compressible vertical wavenumber})
are the clear choice and lead to a final dispersion relationship of
\begin{equation}
  \omega_{RG}^2 \left[ m^2 + k^2 + \frac{1}{4 H^2}\right] = m^2 \gamma N^2,
  \label{eq:RG dispersion relationship}
\end{equation}
which is the same as equation~(\ref{eq:compressible dispersion GW})
except for the factor of $\gamma$ multiplying $N^2$.

While the functional form of the frequencies given in
equation~(\ref{eq:RG dispersion relationship}) are correct up to a
factor of $\sqrt{\gamma}\approx 1.29$ for $\gamma=5/3$, and while the
vertical structure of the eigenfunction matches with the fully
compressible case, we note that this is a special case brought about
by $\vec{\del}\ln T_0=0$ in an isothermal atmosphere.  In more general
atmospheres, an extra term would exist in equation~(\ref{eq:RG wave momentum})
of the form $\pomega \vec{\del}\ln T_0$ and we would be faced by the same
problems with energy conservation and growth that we found in
Section~\ref{sec:ANS waves} for the ANS equations.  We will see this
in Section~\ref{sec:Anelastic-I Energy Balance}.

We summarize the properties of gravity waves for all four systems of
equations in an isothermal atmosphere of infinite extent in
Table~\ref{table:dispersion relationships} and plot them for waves
with $kH=1/10$ in Figure~\ref{fig:omega infinite atm}.
In the low-frequency regime (Figure~\ref{fig:omega infinite atm}$a$),
the gravity wave branch of the exact solutions to the Euler equations
(labelled FC, and given by eq~\ref{eq:omega Navier-Stokes infinite isothermal full dispersion})
matches the dispersion relationship of the LBR equations closely, while the
frequencies of gravity waves in the ANS and RG equations are too large.
As the horizontal wavenumber $m$ increases, the LBR dispersion
relationship begins to diverge from the exact results.  At still
larger wavenumber $m$, both the ANS and LBR dispersion relationships
return to agreement with the exact Euler solutions
(Figure~\ref{fig:omega infinite atm}$b$).  At all wavenumbers, 
the frequencies from the RG equations are a factor of $\sqrt{\gamma}$
larger than those obtained from the LBR equations and thus exceed the 
Brunt-V\"ais\"al\"a frequency $N$ at large $m$.  Higher order radial modes
show similar behavior, though the relative differences between the
LBR and FC dispersion relationships decreases as $k$ increases.

\begin{deluxetable}{ccccccccccccc}
\tabletypesize{\footnotesize}
 \tablecolumns{3}
 \tablewidth{0pt}  
 \tablecaption{Infinite isothermal atmosphere
 \label{table:dispersion relationships}}
 \tablehead{\colhead{System}  &  
   \colhead{$\omega^2=$} &
   \colhead{eq} 
}
\startdata
FC &
$ \left[ m^2 + k^2 + \frac{1}{4 H^2}\right]^{-1} m^2 N^2$
& (\ref{eq:compressible dispersion GW}) \\[2mm]
ANS &
 $\Big[m^2+ k^2 +\frac{1}{4 \gamma^2 H^2}\Big]^{-1} m^2 N^2$
& (\ref{eq:ASH dispersion relationship real omega}) \\[2mm]
ANS-I &
 $\Big[m^2+ k^2 +\frac{1}{4 H^2}\frac{2 - \gamma}{\gamma} + i k g \left(\gamma - 1\right)\Big]^{-1} m^2 N^2$
& (\ref{eq:ASH dispersion relationship}) \\[2mm]
LBR & 
$\left[m^2 + k^2 + \frac{1}{4 H^2}\right]^{-1} m^2 N^2$
& (\ref{eq:LBR dispersion relationship}) \\[2mm]
RG & 
$\left[m^2 + k^2 + \frac{1}{4 H^2}\right]^{-1} m^2 \gamma N^2$
& (\ref{eq:RG dispersion relationship})
\enddata
\tablecomments{Horizontal wavenumbers $m$ and vertical wave
 numbers $k$ are real quantities.  In all systems of equations
 except ANS, we have taken the radial eigenfunctions corresponding
 to equations~(\ref{eq:compressible vertical eigenfunction}--\ref{eq:compressible vertical wavenumber})
 which remain finite in the far-field limit.  In the ANS equations,
 we instead take eigenfunctions~(\ref{eq:ASH vertical wavenumber real omega}--\ref{eq:ASH vertical eigenfunction real omega}),
 which leads to real $\omega$ but divergent behavior in the
 far-field limit.  For the FC equations we here show only
 $\omega^2_{GW}$ (the low-frequency limit);  the full Euler dispersion relationship is
 given in equation~(\ref{eq:omega Navier-Stokes infinite isothermal full dispersion}).}
\end{deluxetable}

\section{Conservation of Energy and Pseudo-Energy}
\label{sec:Energy}
\label{sec:self-adjointness}

An important theme of this paper revolves around energy budgets in
different approximations to the full Euler equations.  The curious
discrepancies found in isothermal atmospheres in 
Section~\ref{sec:isothermal atmosphere} hint at deeper
issues in these approximated equation sets. In this section, 
we find that those issues are associated with energy conservation and
its violation.  Here we consider general atmospheres, with the
isothermal atmospheres of Section~\ref{sec:isothermal atmosphere}
being a subset of these results.  For each set of equations,
beginning with the full Euler equations and proceeding with each
anelastic equation set in turn, we derive the energy conservation
properties for arbitrary nonlinear motions.  We then consider the
energy conserving properties of linearized motions 
including wave-like perturbations.  We find that some equation sets 
(FC and LBR) conserve energy and behave as expected.  We find that the
other anelastic equation sets  (ANS and RG) do not conserve energy and
instead conserve a stratification-weighted pseudo-energy, which leads
to some surprising and paradoxical results for wave-like motions.  The
key results of this section are summarized in 
Table~\ref{table:energies and pseudo energies}.

\begin{deluxetable}{crcccc}
 \tabletypesize{\footnotesize}
  \tablecolumns{6}
  \tablewidth{0pt}  
  \tablecaption{Energies and pseudo-energies
  \label{table:energies and pseudo energies}}
  \tablehead{\colhead{System}  &  
    \colhead{weight} &
    \colhead{eq} &
    \colhead{IA} &
    \colhead{RZ} &
    \colhead{CZ}
}
 \startdata
FC &
$ \rho_0 $
& (\ref{eq:linear Euler energy}) 
& Y & Y & Y \\
 ANS &
$ \exp(-S_0/c_{p}) \rho_0$
& (\ref{eq:Anelastic-I weighted-self-adjoint}) 
& N & N & Y \\
LBR & 
$ \rho_0 $
& (\ref{eq:Anelastic-II self-adjoint}) 
& Y & Y & Y \\
 RG & 
$ T_0^{-1}\rho_0 $
& (\ref{eq:Anelastic-III weighted-self-adjoint}) 
& Y & N & N 
\enddata
\tablecomments{Weight required for self-adjointness and hence energy
  or pseudo-energy conservation in each system of equations, with
  reference to where the conservation properties are shown in the text.  Systems with
  weights other than $\rho_0$ will not always conserve energy.
  Included are qualitative estimates of whether each set of equations 
  is likely to conserve energy in an isothermal atmosphere
  (IA), in the stably stratified solar radiative zone (RZ), and in the
  nearly adiabatically stratified solar convection
  zone (CZ).
}
\end{deluxetable}

\subsection{Euler Energy Balance \label{sec:Euler Energy Balance}}

We begin by considering the Euler equations.
The main results of this subsection are well known in the literature
\citep[e.g.,][]{Lighthill_1978_Waves_in_Fluids}.   Namely, in the
fully compressible Euler equations, energy is conserved by wavelike
motions and the temporal frequencies $\omega$ are purely real.
However, for the purposes of comparison with anelastic models,  we
note that the fully nonlinear equations~(\ref{eq:Euler continuity})--(\ref{eq:Euler entropy}) 
contain a statement of conservation of energy.  Contracting
\eq{eq:Euler momentum} with velocity $\vec{u}$ and assuming that gravity is given
by a potential function gives  
\begin{eqnarray}
&&\pd{E}{t} + \div \left[ \, \vec{u} \, \right(E + P\left)\, \right] \ = \ 0, \label{Euler Energy}
\end{eqnarray}
where
\begin{eqnarray} 
&& E \ = \ \frac{\rho |\vec{u} |^{2}}{2} + \frac{P}{\gamma -1} + \rho\,  \Phi,\\
&& \vec{g}  \ = \ - \grad \Phi, 
\end{eqnarray}
with $\Phi$ the gravitational potential.  The fully compressible Euler
equations conserve energy for arbitrary (nonlinear) motions.

For the linearized version of the Euler equations, we may go a step
further.  For a system in hydrostatic balance 
(eq.~\ref{eq:hydrostatic balance}), we write \eq{eq:linear Euler} in
terms perturbed pressure $P_1$ and entropy $S_1$ as
\begin{eqnarray}
\label{eq:linear Euler 2} 
\rho_{0} \frac{\partial^{2}\vec{\xi}}{\partial t^{2}} = 
- \grad P_{1} 
+ \frac{P_{1}}{\gamma P_{0}} \grad P_{0}
+ \frac{g^2\rho_{0}}{N^{2}}\frac{S_{1}}{c_P}\frac{ \grad{S_{0}}}{c_P}.
\end{eqnarray}
We introduce an arbitrary vector $\vec{\xi^\prime}$ that is related to
the displacement vector $\vec{\xi}$ (eq~\ref{eq:xi}) and guided 
by equations (\ref{eq:frozen-in pressure}) and (\ref{eq:frozen-in entropy}) 
define $P_1^\prime$ and $S_1^\prime$ as
\begin{eqnarray}
  P_1^\prime&=& -\vec{\xi^\prime}\cdot\vec{\del}  P_0 -  \gamma P_0 \vec{\del}\cdot\vec{\xi^\prime}, 
  \label{eq:frozen-in pressure prime}\\
  S_1^\prime &=& -\vec{\xi^\prime}\cdot\vec{\del} S_0.
  \label{eq:frozen-in entropy prime}
\end{eqnarray}
Contracting \eq{eq:linear Euler 2} with arbitrary $\vec{\xi^\prime}$ and using
equations~(\ref{eq:frozen-in pressure prime}) and (\ref{eq:frozen-in entropy prime}) 
gives
\begin{eqnarray}
\label{eq:Euler self-adjoint}
\rho_{0}\, \vec{\xi^{\prime}} \dot \frac{\partial^{2}\vec{\xi}}{\partial t^{2}} 
+ \frac{P_{1}}{\gamma P_{0}} P_{1}^{\prime} 
+ \frac{g^{2}\rho_{0}}{N^{2}}\frac{S_{1}}{c_P}\frac{S_{1}^\prime}{c_P} 
+ \div \left( \vec{\xi^{\prime}} P_{1}\right) = 0. \nonumber\\
\end{eqnarray}

We may derive a number of different results from \eq{eq:Euler self-adjoint}.  
First we consider velocity perturbations and take $\vec{\xi^{\prime}} = \partial_{t} \vec{\xi}=\vec{u}$
(thus $S_1^\prime=\partial_t S_1$ and $P_1^\prime = \partial_t P_1$).
This choice gives the local conservation of energy for linear perturbations 
\begin{eqnarray}
\label{eq:linear Euler energy} \pd{}{t}\left(\frac{\rho_{0}|\vec{u}|^{2}}{2} + \frac{g^{2} \rho_{0}}{2 N^{2}}\left(\frac{S_{1}}{c_{p}}\right)^{2} + \frac{P_{1}^{2}}{2 \gamma P_{0}} \right) + \div \left(\vec{u} \,P_{1} \right) = 0.\nonumber\\
\end{eqnarray}
Integrating \eq{eq:linear Euler energy} over a volume $V$ with $(\vec{u}\cdot\vec{\hat{n}})\,P_{1}= 0$ on the boundary $\partial V $, gives
\begin{eqnarray}
\pd{}{t}\left( K + U \right) = 0,
\end{eqnarray}
where the kinetic and potential energies are given respectively by
\begin{eqnarray}
&& K = \frac{1}{2} \int_{V}\rho_{0}|\vec{u}|^{2}\dV, \\ 
&& U = \frac{1}{2} \int_{V}\left( \frac{g^{2} \rho_{0}}{N^{2}}\left(\frac{S_{1}}{c_{p}}\right)^{2} + \frac{P_{1}^{2}}{\gamma P_{0}} \right)\dV.
\end{eqnarray}
Linear perturbations also conserve energy in the fully compressible Euler equations.

Choosing instead that $\vec{\xi}^{\prime} = \vec{\xi}$ 
(with this choice, $S_1^\prime= S_1$ and $P_1^\prime = P_1$) in \eq{eq:Euler self-adjoint}, 
integrating over volume $V$ with $(\vec{u}\cdot\vec{\hat{n}})\,P_{1}= 0$ on  boundary 
$\partial V $, and averaging over time gives a version of energy equipartition
for linear perturbations, where the time average of  the kinetic
energy equals the time average of the potential energy.  

Rather than considering energy conservation, we now consider the
time-dependence of linearized displacements
\begin{eqnarray}
\vec{\xi} = \vec{\hat{\xi}}e^{i \omega t}, \quad \vec{\xi^{\prime}} =\vec{\hat{\xi}^{*}}e^{-i \omega t},
\label{eq:omega displacement}
\end{eqnarray} 
where $\vec{\hat{\xi}^{*}}$ represents the complex conjugate of
displacement $\vec{\hat{\xi}}$.
Here, $S_1^\prime$ gives the characteristic entropy perturbation
$S_1$ associated with a displacement of amplitude $\vec{\hat{\xi}}$ and
likewise with  $P_1^\prime$ and pressure perturbation $P_1$.  
Integrating over the same volume, $V$, gives  
\begin{eqnarray}
\label{eq:real-omega} \omega^{2} \int_{V}\rho_{0}|\vec{\xi}|^{2}\dV  
- \int_{V}\left( \frac{g^{2} \rho_{0}}{N^{2}} \frac{|S_{1}|^{2}}{c_{p}^{2}} + \frac{|P_{1}|^{2}}{\gamma P_{0}} \right)\dV  = 0. \nonumber\\
\end{eqnarray}
All of the integrals in \eq{eq:real-omega} are strictly real and
positive definite, which implies that the squared temporal frequencies must also be  real
\begin{eqnarray}
\label{eq:real omega squared} \Im\left(\omega^{2}\right) = 0.
\end{eqnarray} 
\Eq{eq:real omega squared} states that while instability may or may
not exist, the system must transition from purely oscillating
($N^{2}>0$) to purely growing behavior ($N^{2}<0$).  Neither growing
nor damped waves exist in the fully compressible Euler equations.  

Lastly, one may show that displacements with different frequencies are orthogonal with respect to the energy inner product,
\begin{equation}
\label{eq:energy inner product}\boldsymbol{\big<} \vec{\xi^{\prime} },  \vec{\xi} \boldsymbol{\big>} \equiv \int_{V} \rho_{0} \vec{\xi^{\prime}}\dot\vec{\xi} \dV  = \delta_{\omega^{\prime}, \omega},
\end{equation}
where $\delta_{\omega^{\prime}, \omega}$ is here the Kronecker delta.
Together, \eq{eq:energy inner product} and 
\eq{eq:Euler self-adjoint} imply that the right-hand side of the linear
perturbation \eq{eq:linear Euler 2} is self-adjoint with respect to
this energy inner product.  Therefore, the condition 
\eq{eq:real omega squared} unlimitedly stems from \emph{both} a particular
dynamical equation, and an appropriate inner product.  If 
\eq{eq:energy inner product} is altered, which amounts to a different
spatial weighting of the solutions, then \eq{eq:real omega squared}
may not hold, and the time dependence of the
solution may acquire spurious growth or decay.  

The above four results that derive from integrating over the volume
$V$ hinge on the condition that
\begin{eqnarray}
\label{boundary condition} 
(\vec{u}\dot\vec{\hat{n}}) \,P_{1}= 0
\label{eq:impenetrable BC}
\end{eqnarray}
on the boundary of $V$ with $\vec{\hat{n}}$ the unit normal vector. 
This condition is not a mere technical triviality, as
\eq{boundary condition} causes the divergence term in \eq{eq:Euler
  self-adjoint} to vanish.  If $V$ is a bounded domain, or is periodic
in the horizontal direction and bounded in the vertical direction,
then we may easily satisfy \eq{boundary condition} by requiring
$(\vec{u}\dot\vec{\hat{n}})= 0$ by itself (e.g., impenetrable
boundaries).  For the travelling waves we considered in
Section~\ref{sec:isothermal atmosphere} the product
$(\vec{u}\dot\vec{\hat{n}}) \,P_{1}$ is itself periodic and integrates
to zero, since $|\vec{u}| \sim \rho_{0}^{-1/2}$, and 
$|P_{1}| \sim \rho_{0}^{1/2}$ for large and small atmospheric heights.
As we will see in the following subsections,
the far-field behavior of travelling waves controls the stability
properties of different anelastic models.

\subsection{ANS Energy Balance \label{sec:Anelastic-I Energy Balance} }
\label{sec:ANS energy}

For comparison with the total energy \eq{Euler Energy} for the Euler
system, we now derive an equivalent energy balance for the anelastic
models, beginning with the ANS equations.  
Contracting \eq{eq:ASH momentum pomega} with  $\vec{u}$ and using
the anelastic continuity \eq{eq:anelastic continuity} gives in basic form 
\begin{equation}
\label{eq:ASH energy} 
\pd{K}{t} + \div \left[\,\vec{u} (K + \rho_{0} \varpi )\,\right] +  \rho_{0}\vec{u} \dot \vec{g} \, \frac{ S_{1}}{c_{p}}  = \\  \varpi \rho_{0} \vec{u}  \dot \grad\left( \frac{S_{0}}{c_{p}}\right)  
\end{equation}
with kinetic energy density $K = \rho_{0} |\vec{u} |^{2}/2$.  
Using the relationship~\ref{eq:conservative buoyancy work I} in Appendix~\ref{sec:appendix conservative buoyancy},
we put the left hand side of equation~(\ref{eq:ASH energy}) into conservative form
\begin{eqnarray}
\label{eq:ASH energy cons form} \pd{E}{t} + \div \left[\,\vec{u} (E + \rho_{0} \hat{\varpi} )\,\right]  \ = \  \varpi \rho_{0} \vec{u}  \dot \grad\left( \frac{S_{0}}{c_{p}}\right) ,  
\end{eqnarray}
where $E$ and $\hat{\pomega}$ are given by
equations~(\ref{eq:anelastic energy E I}) and (\ref{eq:pomega hat}) respectively.
We cannot however transform the right-hand side  into conservative form unless 
\begin{eqnarray}
\label{ASH conservation condition}\lim_{\tau \rightarrow \infty} \frac{1}{\tau}\int_{0}^{\tau} \int_{V} \pomega \rho_{0} \vec{u} \dot \grad{S_{0}} \dV \dt = 0.
\end{eqnarray}
This condition is not true in general and, simply stated, arbitrary (nonlinear)
motions in the ANS equations do not conserve energy.  Condition~(\ref{ASH conservation condition})
is satisfied for adiabatically-stratified atmospheres, where $\grad{S_{0}} =0$, and
in those systems the ANS equations do conserve energy.
 
We turn now to linearized motions to learn more about the strange
behavior found in Section~\ref{sec:ANS waves}  
by considering the equivalent of \eq{eq:Euler self-adjoint}
for the ANS model equations.  Contracting the linear momentum equation
with an arbitrary $\vec{\xi^{\prime}}$, but here satisfying 
$\div \left(\rho_{0} \vec{\xi^{\prime}}\right) = 0$ (again,
$\vec{\xi^\prime}$ could be either $\vec{\xi}$ or $\partial_t \vec{\xi}=\vec{u}$), produces 
\begin{equation}
\label{eq:Anelastic-I non-self-adjoint}
\rho_{0}\left( \vec{\xi^{\prime}} \dot \frac{\partial^{2}\vec{\xi}}{\partial t^{2}} +  
\frac{g^{2}}{N^{2}} \frac{S_{1}^{\prime} S_{1} }{c_{p}^{2} }\right)+ \div \left(\rho_{0} \,\vec{\xi^{\prime}} \pomega \right) = 
\pomega \rho_{0} \, \vec{\xi^{\prime}} \dot \grad \left(\frac{S_{0}}{c_{p} } \right).
\end{equation}
If we integrate \eq{eq:Anelastic-I non-self-adjoint} over the entire
volume, $V$, then the right-hand side refuses to vanish: even linearized
motions do not conserve energy in the ANS equations. 

The non-vanishing right-hand side of \eq{eq:Anelastic-I non-self-adjoint}  
would also appear to imply that the squared frequencies $\omega^2$ are
not strictly real.  On the surface, the asymmetric nature of 
\eq{eq:Anelastic-I non-self-adjoint} would appear to imply
non-self-adjointness of the linear equations and hence spuriously growing modes.
This is consistent with what we found for our analysis in an infinite
isothermal atmosphere (Sec.~\ref{sec:ANS waves}); as we found there, a
correction to the spatial structure counteracts this effect and
regains real eigenvalues for the linear equations at the cost of modes
which grow in spatial height. 
In the literature of anelastic simulations however, no mention appears
of these spuriously growing gravity waves, and a paradox seems
apparent \citep[e.g.,][]{Rogers&Glatzmaier_2005_ApJ, Brun_et_al_2011}.   

The paradox of spurious growth is remedied by the following transformation of \eq{eq:Anelastic-I non-self-adjoint},
\begin{eqnarray}
\label{eq:Anelastic-I weighted-self-adjoint}
\hat{\rho}_{0}\left( \vec{\xi^{\prime}} \dot \frac{\partial^{2}\vec{\xi}}{\partial t^{2}} +  \frac{g^{2}}{N^{2}} \frac{S_{1}^{\prime} S_{1} }{c_{p}^{2} }\right) + \div \left(\hat{\rho}_{0} \,\vec{\xi^{\prime}} \pomega \right) = 0,
\end{eqnarray}
where we define the scaled pseudo-density
\begin{eqnarray}
\hat{\rho}_{0} = \rho_{0}\, e^{-S_{0}/c_{p}},
\label{eq:scaled pseudo-density}
\end{eqnarray}
which reduces to the actual background density in the case of adiabatic stratification.  

Though energy is not conserved for nonlinear dynamics, nor for linear
waves, \eq{eq:Anelastic-I weighted-self-adjoint} implies that the
following pseudo-energy is conserved  
\begin{eqnarray}
\hat{E} = \frac{1}{2}\int_{V}\hat{\rho}_{0}\left[ |\vec{u}|^{2} + \frac{g^{2}}{N^{2}}\left(\frac{S_{1}}{c_{p}}\right)^{2}\right]\dV,
\label{eq:ASH pseudo energy}
\end{eqnarray}
i.e., $\partial_{t} \hat{E} = 0$, for \textsl{linear} perturbations.
If the perturbations are \textsl{nonlinear} then the rescaling of the
density fails since the advection of kinetic energy is not an exact
divergence in terms of this pseudo-density.   

As in the compressible case (eq.~\ref{eq:real-omega}), one may use 
\eqs{eq:omega displacement}{eq:Anelastic-I weighted-self-adjoint}  to show that  
\begin{multline}
\label{eq:ASH real omega squared} 
\omega^{2}\int_{V}\hat{\rho}_{0} |\vec{\xi}|^{2} \dV =  
\int_{V} \hat{\rho}_{0} \frac{g^{2}}{N^{2}}\left|\frac{S_{1}}{c_{p}}\right|^{2}\dV = \\
\int_{V} \hat{\rho}_{0} \frac{N^{2}}{g^{2}} |\vec{\xi}\dot \vec{g} |^{2} \dV,
\end{multline}
which implies that $\Im \left(\omega^{2}\right) = 0 $ even if energy
is not conserved.    This indicates that the conservation of a
pseudo-energy resolves the paradox of spurious growth and leads to
purely real squared temporal frequencies $\omega^2$.  We believe that this
explains why this phenomena of pseudo-energy conservation and energy
violation has been previously missed in the literature. 

\Eqs{eq:Anelastic-I weighted-self-adjoint}{eq:ASH real omega squared} 
imply that the linearized ANS equations are
self-adjoint under the pseudo-energy inner product, and that
eigenfunctions with different frequency are orthogonal with respect to
this pseudo-density weighted norm  
\begin{eqnarray}
\label{eq:pseudo-energy inner product}\boldsymbol{\big<} \vec{\xi^{\prime} },  \vec{\xi} \boldsymbol{\big>}_{\hat{\rho}_{0}} \equiv \int_{V} \hat{\rho}_{0} \vec{\xi^{\prime}}\dot\vec{\xi} \dV  = \delta_{\omega^{\prime}, \omega}.
\end{eqnarray}
The difference between \eqs{eq:energy inner product}{eq:pseudo-energy inner product} 
imply that external forcings and initial conditions project onto
different frequencies and basis vectors differently in the ANS
equations than in the FC equations. 
In particular, the eigenfunctions of pseudo-energy-conserving waves in
the ANS equations are different than the eigenfunctions given by
energy-conserving motions (e.g., the FC equations).  
In strongly stably-stratified atmospheres, these differences may be dramatic,
as we will encounter in Section~\ref{sec:bounded geometries}  

\subsection{LBR Energy Balance \label{sec:Anelastic-II Energy Balance} }

Unlike the ANS equations, the LBR equations show no problems with
energy conservation.  Contracting \eq{eq:LBR momentum pomega} with
$\vec{u}$ and using the anelastic continuity equation~(\ref{eq:anelastic continuity}) 
gives in basic form 
\begin{eqnarray}
\label{eq:LBR energy-0} \pd{K}{t} + \div \left[\,\vec{u} (K + \rho_{0} \varpi )\,\right] +  \rho_{0}\vec{u} \dot \vec{g} \, \frac{ S_{1}}{c_{p}} =   0.  
\end{eqnarray}
Using the relationship~\ref{eq:conservative buoyancy work I} in Appendix~\ref{sec:appendix conservative buoyancy},
we put equation~(\ref{eq:LBR energy-0}) into conservative form
\begin{eqnarray}
\label{eq:LBR energy} \pd{E}{t} + \div \left[\,\vec{u} (E + \rho_{0} \hat{\varpi} )\,\right]  \ = \  0,  
\end{eqnarray}
where $E$ and $\hat{\pomega}$ are given by
equations~(\ref{eq:anelastic energy E I}) and (\ref{eq:pomega hat}) respectively.
If we integrate this over a bounded volume $V$ (where as in
eq~\ref{boundary condition}, $\vec{u} \cdot \vec{\hat{n}}=0$)
then the divergence terms vanish and arbitrary (nonlinear) motions in
the LBR equations obey an energy conservation law. 

For linear perturbations and for nonlinear perturbations in certain
atmospheres (including adiabatic and isothermal atmospheres), the LBR equations conserve an alternative total energy
\begin{eqnarray}
\label{eq:alternative total energy} \tilde{E} \equiv \frac{1}{2}\int_{V} \rho_{0} \left( |\vec{u}|^{2} + \frac{1}{c_{p}}\frac{d\Phi}{d S_{0}} \,S_{1}^{2}  \right) \dV,
\end{eqnarray} 
i.e. $\partial_t \tilde{E} =0$, as detailed in Appendix~\ref{sec:appendix conservative buoyancy}.

For linear perturbations we may furthermore write 
\begin{eqnarray}
\label{eq:Anelastic-II self-adjoint}
\rho_{0}\left( \vec{\xi^{\prime}} \dot \frac{\partial^{2}\vec{\xi}}{\partial t^{2}} +  \frac{g^{2}}{N^{2}} \frac{S_{1}^{\prime} S_{1} }{c_{p}^{2} }\right) + \div \left(\rho_{0} \,\vec{\xi^{\prime}} \pomega \right) = 0.
\end{eqnarray}
This implies self-adjointness of system under the energy inner
product, and also that
\begin{multline}
\label{eq:LBR real omega squared} 
\omega^{2}\int_{V}\rho_{0} |\vec{\xi}|^{2} \dV = 
\int_{V} \rho_{0} \frac{g^{2}}{N^{2}}\left|\frac{S_{1}}{c_{p}}\right|^{2}\dV =\\
 \int_{V} \rho_{0} \frac{N^{2}}{g^{2}} |\vec{\xi}\dot \vec{g} |^{2} \dV,
\end{multline}
whence it follows that $\Im\left(\omega^{2}\right) = 0 $.  Linear
motions conserve energy in the LBR equations and wavelike motions 
have real squared temporal frequencies $\omega^2$.

\subsection{RG Energy Balance \label{sec:Anelastic-III Energy Balance} }

For the RG equations, using similar transformations as in
Sections~\ref{sec:Euler Energy Balance}--\ref{sec:Anelastic-II Energy Balance}, 
we obtain the following nonlinear energy balance 
\begin{eqnarray}
\label{eq:RG energy} \pd{E}{t} + \div \left[\,\vec{u} (E + \rho_{0} \hat{\varpi} )\,\right]  \ = \  
\varpi \rho_{0} \vec{u}  \dot \grad \ln T_{0},
\end{eqnarray}
where
\begin{eqnarray}
&&E = \rho_{0} \left( \frac{|\vec{u}|^{2}}{2} - \gamma \Phi \frac{T_{1}}{T_{0}}\right),\\
&& \hat{\pomega} =  \pomega - \frac{\gamma}{c_{p}}\int_{a}^{r} \Phi(r) \, \mathrm{d}S_{0}(r).
\end{eqnarray}
It is not possible to cast the right-hand side of \eq{eq:RG energy} into conservative form
except in the specialized case of isothermal atmospheres where $\grad \ln T_0= 0$.
This contrasts with the ANS equations, which can only be written in
conservative form in adiabatic atmospheres.
Thus the RG equations do not conserve energy for either anelastic
convection or gravity wave dynamics in arbitrary atmospheres.

Linear perturbations to these equations do nevertheless obey a
pseudo-density weighted self-adjointness 
\begin{eqnarray}
\label{eq:Anelastic-III weighted-self-adjoint}
\hat{\rho}_{0}\left( \vec{\xi^{\prime}} \dot \frac{\partial^{2}\vec{\xi}}{\partial t^{2}} +  
\frac{g^{2}}{\gamma N^{2}} \frac{T_{1}^{\prime}\, T_{1} }{T_{0}^{2} }\right) + 
\div \left(\hat{\rho}_{0} \,\vec{\xi^{\prime}} \pomega \right) = 0,\nonumber\\
\end{eqnarray}
where
\begin{eqnarray}
\frac{ T_{1} }{ T_{0} } = - \gamma \vec{\xi}\dot \grad \left( \frac{ S_{0}}{c_{p}} \right),
\end{eqnarray}
and the pseudo-density becomes
\begin{eqnarray}
\hat{\rho}_{0} =  \frac{\rho_{0}}{T_{0}}.
\label{eq:RG pseudo density}
\end{eqnarray}
Furthermore, as in Sections~\ref{sec:Euler Energy Balance}--\ref{sec:Anelastic-II Energy Balance}, 
we find that
\begin{multline}
\label{eq:RG real omega squared} 
\omega^{2}\int_{V}\hat{\rho}_{0} |\vec{\xi}|^{2} \dV= \int_{V} \hat{\rho}_{0} \frac{g^{2}}{\gamma N^{2}}\left|\frac{T_{1}}{T_{0}}\right|^{2}\dV =\\
 \gamma \int_{V} \hat{\rho}_{0} \frac{N^{2}}{g^{2}} |\vec{\xi}\dot \vec{g} |^{2} \dV.
\end{multline}
Equation~(\ref{eq:RG real omega squared}) implies that $\Im\left( \omega^{2} \right) = 0 $ even if
energy is not conserved.  As in the ANS equations, waves in the RG
equations have real squared temporal frequencies $\omega^2$ in volumes
where $\hat{\rho}_{0} \,(\vec{\xi}\cdot\vec{\hat{n}}) \pomega =0$ on
the domain boundaries, but the eigenfunctions and energies are
weighted by pseudo-density~(\ref{eq:RG pseudo density}).

Equation~(\ref{eq:RG real omega squared}) implies that the stability
\emph{boundary} for the fully compressible and other 
anelastic models, $N^{2} = 0$, remains unaltered in the RG equations
in spite of energy non-conservation.   Two problems do however
still remain.  The first is that an extra factor of $\gamma$ appears in
the last integral of \eq{eq:RG real omega squared}.  As we found for
waves in an isothermal atmosphere, this leads to frequencies that are
too high.  The second more serious issue is that
energy is not conserved \textsl{unless} the background atmosphere is
isothermal. In particular, both linear and nonlinear motions within
adiabatically-stratified atmospheres will not conserve energy.

\section{Bounded atmospheres and implementation in spherical systems}
\label{sec:bounded geometries}

We now turn to considering gravity waves in a spherical shell.  
In this geometry divergence at infinity is no longer a problem.
We will find that impenetrable boundary conditions at the top and
bottom of the spherical shell lead to frequencies that are purely real 
(e.g., oscillating waves only, with no spuriously growing modes) but
now the eigenfunctions will be severely distorted in the ANS equations
as compared with the LBR equations.  In a sense, the eigenfunctions
try to diverge to infinity but are constrained by the boundary
conditions.  Analytic eigenfunctions can be found if we consider a simplified
atmosphere with constant gravity $\vec{g} = -g \vec{\hat{r}}$ and
constant Brunt-V\"ais\"al\"a frequency $N$.   In an isothermal atmosphere with
temperature $T_0$ this can be achieved by setting the entropy gradient to 
\begin{equation}
  \del_r S_0 = \frac{\partial S_0}{\partial r} = \frac{g}{T_0}.
  \label{eq:isothermal dsdr}
\end{equation}
The background entropy $S_0$ is found by integration, with the
arbitrary constant set by a reference value within the atmosphere
(here at the base of the domain).   The background pressure and
density are determined by hydrostatic balance and their values at the
reference layer. 
We first derive the analytic solutions and then
compare these solutions with fully nonlinear calculations using two
versions of the anelastic spherical harmonic (ASH) code.

\subsection{Modes in stratified isothermal spherical shells}
\label{sec:isothermal analytic bounded}

We begin by obtaining analytic solutions for the low-Mach number ANS,
LBR, and RG equations.  Full details are given in 
Appendix~\ref{appendix:eigenfunctions}.
Motivated by the properties of the solar radiative zone, we solve for the
eigenvalues of the low-Mach number anelastic equations within a spherical shell
stretching from  $a = 0.50 R_\odot$ to $b = 0.70 R_\odot$ with
$r_\odot$ the solar radius.  This shell has geometric extent
\begin{equation}
  \chi = \frac{a}{b} = 0.717
\end{equation}
and we consider several different values for the scale height $H$ and
number of density scale heights $n_\rho$.  The atmospheric parameters
are reported in Table~\ref{table:atmospheres}.  
The first five such wavenumbers for the ANS and LBR equations are 
presented in Table~\ref{table:numerical vertical wavenumbers} for
several of these atmospheres.

\begin{deluxetable}{ccccccccccccc}
 \tabletypesize{\footnotesize}
  \tablecolumns{7}
  \tablewidth{0pt}  
  \tablecaption{Atmosphere parameters
  \label{table:atmospheres}}
  \tablehead{
    \colhead{$n_\rho$} &
    \colhead{$\Delta S/c_P$} &
    \colhead{$H$}  &  
    \colhead{$H$} &
    \colhead{$T_0$} &
    \colhead{$N$} &
    \colhead{$\tau_\mathrm{BV}$} \\
    \colhead{} &
    \colhead{} &
    \colhead{$Mm$}  &  
    \colhead{$R_\odot$} &
    \colhead{$10^6 K$} & 
    \colhead{$10^{-3} s^{-1}$} &
    \colhead{$s$}
}
 \startdata
 \phn 0.25      & 0.1            & 548\phantom{.} & 0.788\phn & 39.3408\phn & 0.84 & 1185\phantom{.} \\ 
 \phn1.0\phn & 0.4      & 137\phantom{.} & 0.197\phn & \phn 9.83520 & 1.69 & 592.6 \\ 
 \phn2.5\phn & 1\phn & 54.8 & 0.0788                         & \phn 3.93408 & 2.66 & 374.6\\
 \phn5.0\phn & 2\phn & 27.4 & 0.0394                         & \phn 1.96704 & 3.77 & 264.9 \\
 \phn7.5\phn & 3\phn & 18.3 & 0.0263                         & \phn 1.31136 & 4.62 & 216.3 \\
 10.0\phn & 4\phn & 13.7 & 0.0197                               & \phn 0.98352 & 5.34 & 187.4 \\
 12.5\phn & 5\phn & 11.0 & 0.0158                               & \phn 0.78682 & 5.96 & 168.5 
\enddata
\tablecomments{Quoted are the number of density
  scale heights in the domain $n_\rho$, the non-dimensional entropy
  drop across the shell $\Delta S/c_P$, the physical size of the density
  scale height $H$ in megameters and relative to the solar radius, the
  isothermal temperature $T_0$, the constant Brunt-V\"ais\"al\"a
  frequency $N$ and the corresponding timescale
  $\tau_{BV} = 1/N$.  In all models, 
  $r_{\mathrm{bot}} = a = 210Mm \approx 0.30 R_\odot$ and 
  $r_{\mathrm{top}} = b = 485Mm \approx 0.70 R_\odot$, with 
  $\chi = a/b = 0.433$ and with the solar radius $r_\odot = 695Mm$.
  Additional simulations conducted at $\Delta S/c_P=10^{-2}$, $10^{-3}$ and
  $10^{-4}$ are not shown here.}
\end{deluxetable}

\begin{deluxetable}{lcccccccccccc}
\tabletypesize{\footnotesize}
 \tablecolumns{6}
 \tablewidth{0pt}  
 \tablecaption{Isothermal atmosphere solutions
 \label{table:numerical vertical wavenumbers}}
 \tablehead{\colhead{}  &  
   \colhead{$k_1$} &
   \colhead{$k_2$} &
   \colhead{$k_3$} &
   \colhead{$k_4$} &
   \colhead{$k_5$}
}\startdata
\sidehead{$n_\rho=2.5$}
ANS & 9.72953 & 19.0635 & 28.4829 & 37.9246 & 47.3752 \\ 
LBR & 10.1839 & 19.3017 & 28.6431 & 38.045 & 47.4717 \\[1mm]
%

\sidehead{$n_\rho=5$}
ANS & 10.4846 & 19.4637 & 28.7527 & 38.1277 & 47.5379 \\ 
LBR & 12.0834 & 20.3817 & 29.3825 & 38.605 & 47.9216 \\[1mm]

\sidehead{$n_\rho = 7.5$}
LBR & 14.6998 & 22.0671 & 30.5764 & 39.5211 & 48.6627 \\
ANS & 11.6324 & 20.1134 & 29.1969 & 38.4639 & 47.8080 \\
FC-1 & 2.16693 & 15.4867 & 26.2203 & 36.2565 & 46.0510 \\
FC-$\infty$ & 9.49926 & 18.6646 & 27.9660 & 37.3421 & 46.7676 \\[1mm]

\sidehead{$n_\rho=12.5$}
ANS & 14.692 & 22.0618 & 30.5726 & 39.5182 & 48.6602 \\ 
LBR & 20.888 & 26.7604 & 34.1169 & 42.3167 & 50.9577  \\

\enddata
\tablecomments{Radial wavenumbers for gravity waves in selected bounded
  isothermal atmospheres 
  listed in Table~\ref{table:atmospheres}. Quoted are the five
  lowest wavenumbers $k_1$--$k_5$ in each equation set.  In the full
  Euler equations, the radial wavenumber depends on
  spherical harmonic $\ell$; as such, we quote wavenumbers at low and
  high values of $\ell$ (FC-$1$ at $\ell=1$ and FC-$\infty$ at
  $\ell=50$ respectively).  The RG equations have the same
  eigenfunctions and radial wavenumbers as the LBR equations 
  and are not separately quoted.
}
\end{deluxetable}

\begin{figure}
 \begin{center}
    \includegraphics[width=0.9\linewidth]{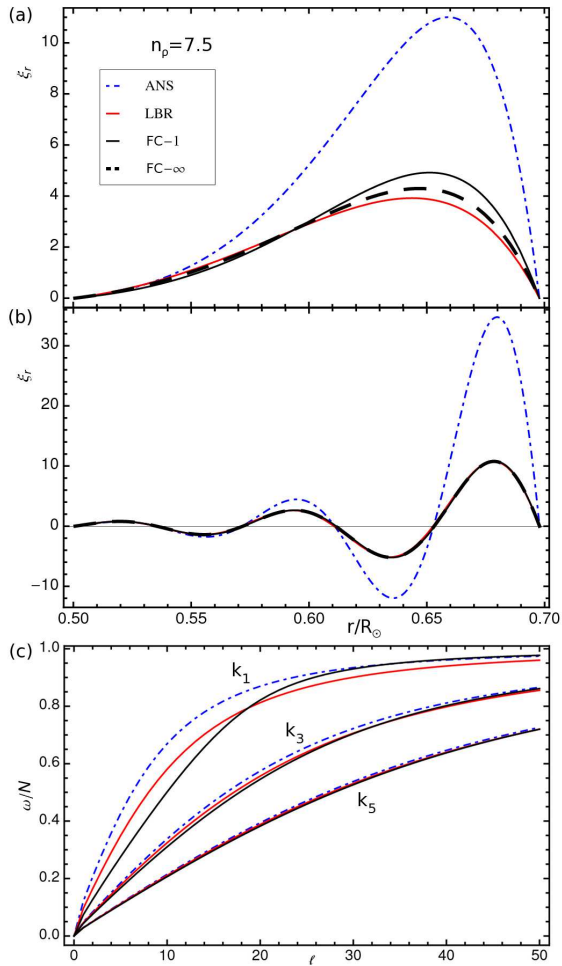}
  \end{center}
\caption{Eigenfunctions for the $n_\rho=7.5$ atmosphere. 
$(a)$ Eigenfunctions for the fundamental $k_1$ mode and 
$(b)$ the higher-order $k_5$ mode.
$(c)$ Dispersion relationship $\omega/N$ for the first, third and
fifth radial modes ($k_1$, $k_3$, $k_5$), with lower-k having higher $\omega$. 
In each plot, the ANS equations are shown in blue (dash-dotted) while
the LBR equations are shown in red (solid).  
The full compressible results are shown in black.  For the
eigenfunctions, the solid line corresponds to $\ell=1$ (FC-$1$) and
the thick dashed line corresponding to $\ell=50$ (FC-$\infty$).
\label{fig:N7.5 eigenfunctions}} 
\end{figure}

We begin by discussing eigenfunctions in the $n_\rho = 7.5$
atmosphere, as this atmosphere will form the primary comparison 
case for the 3-D numerical simulations in
Section~\ref{sec:ASH sims}.  In Figure~\ref{fig:N7.5 eigenfunctions} we
show both the fundamental $k_1$ mode and a higher-order $k_5$ mode.
In addition to the various low-Mach number anelastic eigenfunctions, here we
also overplot eigenfunctions for the fully compressible Euler (FC)
equations; and these require numeric solutions.  
In the full FC equations, the radial eigenfunction depends
on spherical harmonic $\ell$,  whereas in the anelastic equation sets
this coupling disappears.  This effect is most pronounced in the FC
eigenfunctions at low-$\ell$, with the eigenfunctions largely becoming
constant with $\ell$ when $\ell \gg k$.  As such, we plot two FC eigenfunctions,
one at $\ell=1$ (FC-$1$) and one at $\ell=50$ (FC-$\infty$).
 
As is clearly evident in Figure~\ref{fig:N7.5 eigenfunctions}$b$, 
the discrepancies in the ANS eigenfunctions do not diminish at high
radial wavenumbers. This continues to hold true for higher
wavenumbers than we show here.  This is not surprising, as these
discrepancies arise from the energy non-conservation in the ANS
equations, rather than from assumptions about the relative size of the
gravity wavelengths and scale heights in the atmosphere.  
In contrast, at high-$k$, the other equation sets all converge.
The dispersion relationship for odd modes $k_1$, $k_3$ and $k_5$ are shown as
a function of spherical harmonic $\ell$ in Figure~\ref{fig:N7.5 eigenfunctions}$(c)$.
For the $k_1$ mode and at low-$\ell$, the low-Mach number anelastic equations
generally produce higher frequencies than the full compressible Euler
equations.  At higher-$\ell$ all of these frequencies converge to the
Brunt-V\"ais\"al\"a frequency $N$, and the frequencies in the ANS and LBR
equations generally  cross the frequencies of the Euler
equations at some moderate $\ell$.  The frequencies converge much sooner at
high radial order (e.g.,~$k_5$). 
The RG equations are not shown in Figure~\ref{fig:N7.5 eigenfunctions}; 
their frequencies are consistently a factor of
$\sqrt{\gamma}\approx1.3$ larger than the LBR equations. 

\begin{figure*}
\begin{center}
  \includegraphics[width=0.9\linewidth]{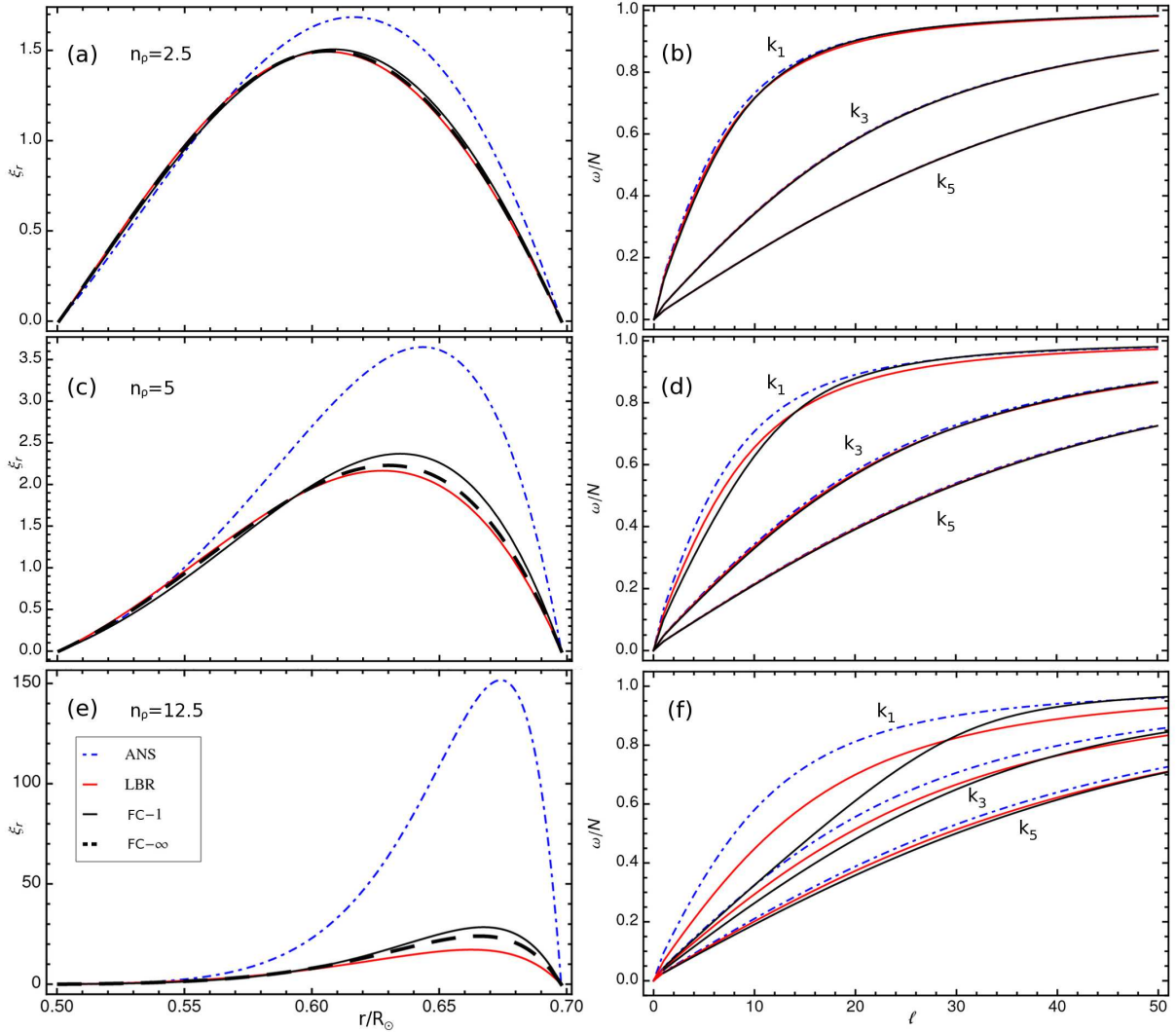}
\end{center}
\caption{Eigenfunctions and dispersion relationships for selected isothermal
  atmospheres.  In $(a,b)$ $n_\rho=2.5$, in $(c,d)$ $n_\rho=5$, and in
  $(e,f)$ $n_\rho=12.5$.  
Eigenfunctions for the fundamental $k_1$ mode are shown in $(a, c, d)$
and dispersion relationships for the $k_1$, $k_3$ and $k_5$ radial
modes are shown in $(b, e, f)$. Labels for lines in all plots are given in $(e)$.
\label{fig:eigenfunctions}} 
\end{figure*}

The eigenfunctions of the fundamental mode $k_1$ are shown in
Figure~\ref{fig:eigenfunctions} for several isothermal atmospheres
from Table~\ref{table:atmospheres}.   With the normalization that we
have chosen (Appendix~\ref{appendix:eigenfunctions}), 
the ANS eigenfunctions are generically larger in amplitude
than the other systems of equations.  This difference is most
pronounced near the top of the domain, and the discrepancies grow as
the amount of stratification grows. 

\subsection{Numerical models with the ASH code}
\label{sec:ASH sims}

We turn now to fully nonlinear 3-D simulations of gravity wave propagation using
the ASH code.  We study gravity waves in ASH using both the standard
ANS equations as well as an implementation of the LBR equations.
In the ASH-ANS equations, the momentum and energy equations are  
\begin{equation}
  \rho_0\left[\frac{\partial \vec{u}}{\partial t} + \vec{u}\cdot\vec{\del}\vec{u}\right] = 
  -\vec{\del}P_1 + \rho_0\frac{P_1}{\gamma P_0}\vec{g} -
  \rho_0\frac{S_1}{c_p}\vec{g} -\vec{\del}\cdot\vec{\scrD},
\label{eq:ASH code momentum}
\end{equation}
\begin{multline}
  \frac{\partial S_1}{\partial t} +\vec{u}\cdot\vec{\del}S_1 =
  -\vec{u}\cdot\vec{\del}S_0 
+ \frac{1}{\rho_0 T_0} \vec{\del}\cdot \left[\kappa \rho_0 T_0 \vec{\del} S_1 \right] \\
+ 2 \frac{\nu}{T_0} \left[e_{i j} e_{i j} - \frac{1}{3} (\vec{\del}\cdot\vec{u})^2\right],
\label{eq:ASH code entropy}
\end{multline}
where the viscous stress tensor is
\begin{equation}
  \vec{\scrD}_{i j} = -2 \rho_0 \nu \left[ e_{i j} - \frac{1}{3}
    (\vec{\del}\cdot\vec{u}) \delta_{i j} \right],
\end{equation}
with $e_{i j}$ the strain rate tensor and $\delta_{i j}$ the
Kronecker delta.   These anelastic equations assume a linearized
equation of state (eq~\ref{eq:linearized EOS}) and the anelastic
constraint (eq~\ref{eq:anelastic continuity}) but are otherwise fully nonlinear.
The ASH-LBR equations are identical except for the momentum equation,
where
\begin{equation}
  \rho_0\left[\frac{\partial \vec{u}}{\partial t} + \vec{u}\cdot\vec{\del}\vec{u}\right] = 
 -\rho_0\vec{\del}\left(\frac{P_1}{\rho_0}\right)
  -\rho_0\frac{S_1}{c_p}\vec{g}
 -\vec{\del}\cdot\vec{\scrD}.
 \label{eq:ASH-LBR momentum}
\end{equation}
All other properties of the simulations are identical.  

We take the geometry and atmosphere used
previously in this section for the background reference state entropy
$S_0$, pressure $P_0$, temperature $T_0$, and density $\rho_0$.  These
quantities vary in radius but do not evolve in time.  
Here we first focus on simulations conducted in an isothermal atmosphere
with $n_\rho =7.5$ and with other parameters given in
Table~\ref{table:atmospheres}.  
In comparison, over the same range of radii in the Sun
$n_{\rho,\odot}\approx 1.8$, while  
$n_{\rho,\odot}\approx 6.6$ over the whole solar radiative zone.   
As such, the results presented here are likely an over-estimate for
comparable effects in the solar interior, but the larger number of
scale heights more clearly emphasizes the differences between the
ASH-ANS and ASH-LBR equations. 
We will return to solar conditions at the end of this section. 

In the pseudo-energy conserving ANS equations, the scaled
pseudo-density (eq~\ref{eq:scaled pseudo-density}) is weighted by the
background entropy $S_0$.  As such, the non-dimensional entropy drop
across the domain
\begin{equation}
  \Delta S/c_P = \frac{1}{c_P}\Big(S_0(r_\mathrm{top}) - S_0(r_\mathrm{bot})\Big),
  \label{eq: delta S}
\end{equation}
or the number of pseudo-density scale heights $n_{\hat{\rho}}$ with
\begin{equation}
  n_{\hat{\rho}} = n_\rho + \Delta S/c_P,
  \label{eq: N pseudo rho}
\end{equation}
are both likely better measurements of how much pseudo-energies differ
from energies in the ANS equations than the number of density scale
heights $n_\rho$ in the atmosphere.  For a non-isothermal atmosphere,
we can use the equation of state (\ref{eq:background EOS}) to obtain
\begin{equation}
  n_{\hat{\rho}} = \frac{2 \gamma -1}{\gamma} n_\rho - \frac{1}{\gamma} n_T ,
\label{eq: N pseudo rho arbitrary}
\end{equation}
where $n_T = \ln{(T_{0,\mathrm{bot}}/T_{0,\mathrm{top}})}$ is the number
of temperature scale heights.  In an isothermal atmosphere with
$\gamma=5/3$, this reduces to $n_{\hat{\rho}} = (7/5) n_\rho$ and the
number of ANS pseudo-density scale heights always exceeds the number of
density scale heights.  
Our isothermal atmosphere with $n_\rho=7.5$ has $\Delta S/c_p =3$ and
$n_{\hat{\rho}} = 10.5$.

The numerical simulations were conducted  in a non-rotating system
with viscosity $\nu = 1 \times 10^{10}\textrm{cm}^2/\textrm{s}$ and entropy
diffusivity $\kappa = 4 \times 10^{10}\textrm{cm}^2/\textrm{s}$ and
with $c_P=3.4\times 10^8~\textrm{ergs}\, \textrm{g}^{-1} \, \textrm{K}^{-1}$.   
In contrast to ASH simulations of stellar convection
\citep[e.g.,][]{Brown_et_al_2008, Miesch_et_al_2008}, 
in these isothermal atmosphere simulations we neglect
radiative diffusion of temperature in the entropy equation and
diffusion of the background entropy gradient $\vec{\del} S_0$, which
in these simulations is set by equation~(\ref{eq:isothermal dsdr}).  
As such, there is no energy flux through the simulation.  The velocity boundary
conditions at the top and bottom of the domain are stress-free and
impenetrable, and the thermal boundaries maintain a constant entropy gradient.
All simulations are conducted with a resolution
of $257 \times 256 \times 512$ ($N_r \times N_\theta \times N_\phi$), with
all functions expanded in Chebyshev polynomials radially and spherical
harmonics horizontally; this leads to a dealiased spectral resolution
of $\ell_\mathrm{max}=170$, which resolves the wave motions studied here.

\begin{figure}
 \begin{center}
   \includegraphics[width=\linewidth]{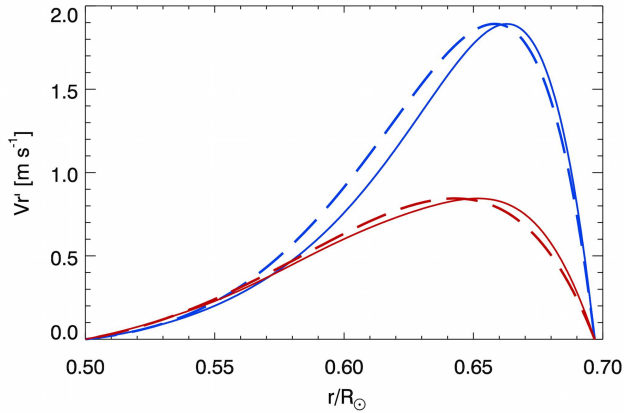}
 \end{center}
\caption{Fluctuating velocities in numerical simulations for the
 $n_\rho=7.5$ isothermal atmosphere.  Shown are
 the rms radial velocities for ASH simulations calculated with ANS
 (blue, solid) and LBR (red, solid) treatments of the momentum
 equation.  The thick dashed lines give the analytic $k_1$ eigenmode
 for each equation set, normalized by the peak velocity realized in
 the ASH simulations. 
 \label{fig:ANS and LBR eigenfunctions}}
\end{figure}

Timestepping errors can have important impacts on the properties of
wave motions.  The ASH code uses a second-order 
Adams-Bashforth/Crank-Nicolson technique for time evolution, which treats diffusive
processes implicitly and advective processes explicitly.  In
our studies here, we found that it is crucial that the advective
interactions between the wave motions and the background reference
state stratification be handled implicitly (on the Crank-Nicolson
side).  In the entropy equation~(\ref{eq:ASH code entropy}), this term is
\begin{displaymath}
  \vec{u}\cdot\vec{\del} S_0.
\end{displaymath}
If these interactions are handled explicitly (via the
Adams-Bashforth portion) then the solutions are sensitive to the size
of the timestep; with sufficiently small timesteps a solution can be
time-evolved correctly, but these timesteps must be nearly an order of
magnitude smaller than are otherwise possible.  Larger timesteps
lead to explicit timestep errors that grow quickly in the solution.
Handling these interactions implicitly, as we do here, leads to much
more stable behavior.   To simplify matters, in these studies we fix
the timestep at slightly less than one third of the 
Brunt-V\"ais\"al\"a timescale $\tau_\mathrm{BV}$ (e.g., $70 s$ in the $n_\rho=7.5$ atmosphere).

At the start of each simulation, random entropy perturbations are introduced
in a band of spherical harmonic $\ell$ ranging from $\ell=1$--$30$, at
all spherical harmonic $m$ values.  The radial perturbation has two
bumps in radius, defined by
\begin{equation}
f(r) = 1-3 x^2 +3x^4 -x^6 +2.5x-2.5x^3,
\label{eq:radial perturbation}
\end{equation}
with scaled radius $x$ given by
\begin{equation}
x = (2 r - r_\mathrm{top} -r_\mathrm{bot})/(r_\mathrm{top}-r_\mathrm{bot}),~x\in [-1,1].
\end{equation}
This radial perturbation does not exactly match the radial
eigenfunction of the gravity waves but rather drives a broad band of
such waves with the largest power in the lowest $k_1$ and $k_2$ modes.  
The initial perturbations lead to flows of roughly $1\ms$ in amplitude,
with Reynolds numbers $Re=u L/\nu$ of about 100.  The
viscous $Q$ of these waves,
\begin{equation}
  Q \equiv \frac{\omega L^2}{\nu} \approx 8.67 \times 10^7
  \label{eq:Q}
\end{equation}
where we have taken $\omega \approx N$ 
(the low-frequency $\ell=1$ fundamental mode has $\omega \approx 0.12 N$) 
and $L = r_\mathrm{top}-r_\mathrm{bot}$,
the depth of our shell.  A thermal $Q$ would be four times smaller.
The $Q$ calculated in equation~(\ref{eq:Q}) is most applicable to our
longest wavelength modes; our shortest wavelength modes with $\ell=30$
would have a $Q$ of about
\begin{equation}
  Q_{30} = \frac{Q}{\ell(\ell+1)} \approx 9.32 \times 10^4,
\end{equation}
which is still quite large.  
Thus, we expect that the gravity waves should only be very weakly
damped by diffusion.

\subsection{Eigenfunctions and violation of energy conservation}
We expect that there will be two clear effects from the choice of ANS or LBR
equations.  The first such effect is that radial eigenfunctions of the
two systems should differ strongly, as discussed in 
Section~\ref{sec:isothermal analytic bounded}.
The radial eigenfunctions for the ASH-ANS and ASH-LBR simulations are
shown in Figure~\ref{fig:ANS and LBR eigenfunctions}.  Plotted on the
same scale and against radius are fluctuating rms radial velocities
$Vr'$ at a time late in the simulations 
(30 days after initiation, or roughly 12,000 $\tau_{BV}$).  
These rms velocities are further time-averaged over roughly 2.5 days
or about 1000 $\tau_{BV}$.

\begin{figure*}[th]
 \begin{center}
    \includegraphics[width=\linewidth]{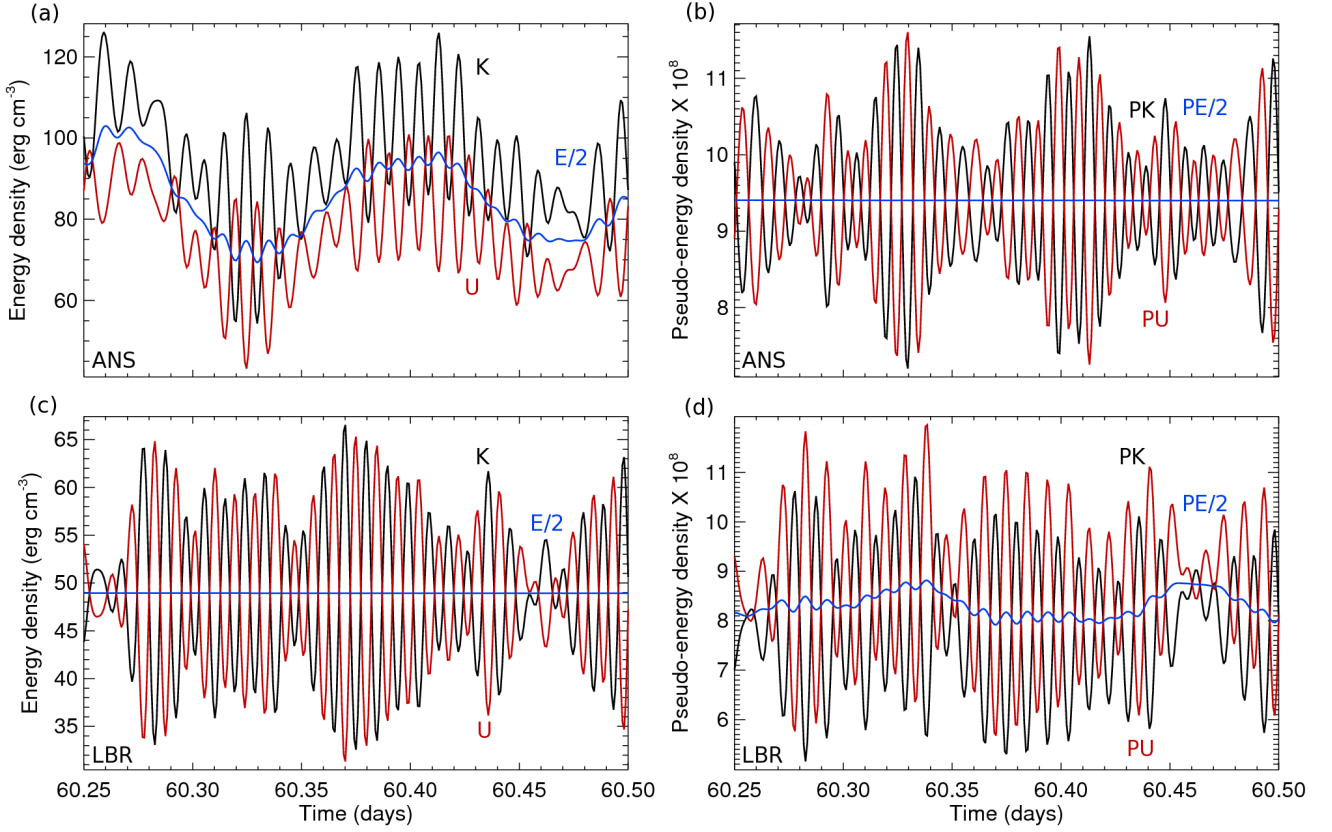}
 \end{center}
\caption{Temporal evolution of energies and pseudo-energies, shown
  over identical intervals for ASH-ANS and ASH-LBR.  $(a)$
  Energy and $(b)$ pseudo-energy in ASH-ANS simulation, with
  definitions as given in equations~(\ref{eq:E}--\ref{eq:PU}). 
  Pseudo-energy densities are here multiplied by $10^8$.
  The total energy E and pseudo energy PE is divided by 2 to highlight the
  fluctuations between K and U or PK and PU respectively.  The ASH-ANS
  equations clearly do not conserve energy E but clearly do conserve
  pseudo-energy PE.  Over the interval shown, $\Delta E \approx 0.101$ while
  $\Delta PE < 10^{-5}$, as defined in equation~(\ref{eq:VE and VP}). 
  $(c)$ Energy and $(d)$ pseudo-energy for ASH-LBR
  simulation.  Energy is clearly conserved in this system, while
  pseudo-energy is not, with $\Delta E < 10^{-6}$ while $\Delta PE
  \approx 0.03$.  
  The temporal interval shown in all plots spans about
  100$\tau_{BV}$ and begins about 24,000$\tau_{BV}$ after the start
  of the simulations.  
  \label{fig:ANS and LBR E and PE}}
\end{figure*}

Overplotted on each simulation is the appropriate radial eigenfunction
corresponding to the gravest $k_1$ mode for the ANS or LBR equations.
Here the eigenfunctions are scaled by the peak rms velocity.  
In both simulations, the rms velocities from the fully nonlinear 3-D
numerical simulations agree very well with the analytic eigenfunctions.  
In the ASH-ANS set of equations, the radial velocities peak more
strongly in the upper portion of the domain, reaching amplitudes 2-4
times larger than the ANS-LBR equations.  The fluctuating velocities differ
strongly, as expected.

The second effect is that, as discussed in Section~\ref{sec:self-adjointness}, the
LBR equations should conserve energy while the ANS equations conserve
a pseudo-energy.  In the simulations we define volume-averaged total energy $E$,
kinetic energy $K$ and potential energy $U$ densities
\begin{eqnarray}
  E &=& K + U, \label{eq:E}\\
  K &=& \frac{1}{V}\int \frac{1}{2} \rho_0 u^2 dV, \\
  U &=& \frac{1}{V}\int \frac{1}{2} \rho_0 g\left(\frac{\partial}{\partial
      r}\frac{S_0}{c_P} \right)^{-1}\left(\frac{S_1}{c_P}\right)^2 dV,
\end{eqnarray}
with fluctuating velocity $u$ and fluctuating entropy $S_1$, and
where the integral is over the full simulation volume $V$ 
(e.g.,~eqn~\ref{eq:alternative total energy}).  
Likewise we define pseudo-energy densities (e.g.,~eqn~\ref{eq:ASH pseudo energy})
\begin{eqnarray}
  PE &=& PK + PU, \\
  PK &=& \frac{1}{V}\int \frac{1}{2} e^{-S_0/c_P}\rho_0 u^2 dV, \\
  PU &=& \frac{1}{V}\int \frac{1}{2} e^{-S_0/c_P}\rho_0 g\left(\frac{\partial}{\partial
      r}\frac{S_0}{c_P} \right)^{-1}\left(\frac{S_1}{c_P}\right)^2 dV. \label{eq:PU}\nonumber\\
\end{eqnarray}
In an isothermal atmosphere with constant Brunt-V\"ais\"al\"a
frequency $N$, $S_0$ is a function of radius and the stratification
term cannot be factored out of the integral.
If thermal and viscous diffusion can be neglected, $E$ should be
conserved in the LBR equations while $PE$ will vary in time.  
Likewise, the ANS equations should conserve total pseudo-energy $PE$
but should fail to conserve total energy $E$. Indeed, this is what we find.

Shown in Figure~\ref{fig:ANS and LBR E and PE} are temporal traces of
energy and pseudo-energy in the ASH-ANS simulation and the simulation
using the ASH-LBR equations.  Here a short interval, spanning about 100$\tau_{BV}$, is shown from a much longer
simulation.   The wave periods are generally longer than $\tau_{BV}$, owing to their long horizontal wavelengths.
The ASH-ANS simulation shows large variations in kinetic and potential
energies K and U and does not conserve total energy E
(Fig.~\ref{fig:ANS and LBR E and PE}$(a)$).  
This simulation does however clearly conserve total pseudo-energy PE 
(Fig.~\ref{fig:ANS and LBR E and PE}$(b)$).   Over much longer
intervals of time, both the total energy and total pseudo-energy decay
dissipatively.  In contrast, the ASH-LBR simulation correctly
conserves energy E (Fig.~\ref{fig:ANS and LBR E and PE}$(c)$) while
the pseudo-energy PE fluctuates in time (Fig.~\ref{fig:ANS and LBR E and PE}$(d)$)

We define the
relative energy variation $\Delta E$ and pseudo-energy variation $\Delta PE$ as
\begin{equation}
  \Delta E = (\delta E)/\langle E \rangle, 
 ~\text{and}~
  \Delta PE = (\delta PE)/\langle PE \rangle,
\label{eq:VE and VP}
\end{equation}
with $\delta$ signifying the standard deviation in time and brackets denoting a
time average over the same period.   Subtracting off the slow diffusive decay, we find that
over a ten-day interval (4000 $\tau_\mathrm{BV}$), $\Delta E \approx
5.6\%$ in the ASH-ANS simulation (during the 
interval shown in Figure~\ref{fig:ANS and LBR E and PE}, 
$\Delta E \approx 10.1\%$).  Over the same interval, $\Delta PE < 0.001\%$.  
In contrast, the ASH-LBR simulation has $\Delta E < 0.0001 \%$ and
$\Delta PE \approx 3\%$ over the same ten-day interval of time. 
We have conducted similar simulations with diffusivities $\nu$ and $\kappa$
ten times larger (e.g., $Re\approx 10$) and find a similar level of
variability, and thus conclude that our results are not strongly
dependent on the level of diffusivity employed.  For the linear waves
considered here, we find that  $\Delta E$ and $\Delta PE$ are
independent of the initial perturbation amplitude.

In these many-wave simulations, the introduced waves span varying
portions of the frequency dispersion relationship, including regions
where $\omega$ depends almost linearly on $\ell$ and regions where it
does not (e.g., Figure~\ref{fig:N7.5 eigenfunctions}$(c)$).
Thus we might expect a collection of these very linear waves to behave as
incoherent oscillators and that the relative energy variations for
many waves might be smaller than those of any individual horizontal
wave.

\begin{figure}[t]
 \begin{center}
   \includegraphics[width=\linewidth]{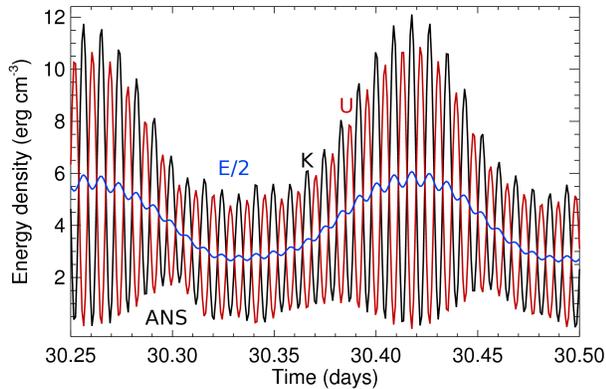}
\end{center}
\caption{Relative energy variation $\Delta E$ in the single-wave
 ASH-ANS solution for the atmosphere with $n_\rho=7.5$ and $\Delta
 S/c_P=3$.  This should be compared with Figure~\ref{fig:ANS and LBR E and PE}$(a)$.
 Generally, the variations in these single wave solutions are about a
 factor of 5 larger than the many-wave solutions.  
 \label{fig:single wave nrho 7.5}}
\end{figure}

This is confirmed by simulations where only a single spherical
harmonic perturbation is initially introduced, as shown by traces of $E$, $K$
and $U$ in Figure~\ref{fig:single wave nrho 7.5} 
for the $n_\rho=7.5$ and $\Delta S/c_P=3$ atmosphere.  In this ASH-ANS
simulation, only $\ell=30$ waves (at all $|m|<\ell$) are initially excited, with the same radial
perturbation as the many-wave simulations (eq~\ref{eq:radial perturbation}).
Hereon, we will refer to these as single-wave solutions.
Comparing Figure~\ref{fig:single wave nrho 7.5} with
the corresponding many-wave solution in Figure~\ref{fig:ANS and LBR E and PE}$(a)$
it is clear that the relative energy variation is significantly greater.
Here, $\Delta E \approx 25\%$ ($\Delta PE$ remains negligible).  
We have studied single-wave solutions with different horizontal
wavelengths, sampling in the range from $\ell=1$ to $\ell=100$ and find that this
level of energy variation is reasonably representative for individual
waves of any horizontal wavelength in this range.  This confirms our
understanding that the phenomenon of energy non-conservation is due to
the level of stratification and does not depend strongly on the
particulars of any single mode (e.g., horizontal or vertical wavelength).

We now turn to considering isothermal atmospheres with differing
levels of stratification, ranging from $n_\rho =0.1$--$12.5$ and
$\Delta S/c_P = 0.1$--$5$ (see Table~\ref{table:atmospheres}). 
The configurations of the simulations are
the same as previously discussed, though at large stratification
($n_\rho \ge 10$) a higher resolution was used, with $N_r = 1025$ 
and a dealiased horizontal resolution of $\ell_\mathrm{max}=340$.

\begin{figure}[t]
  \begin{center}
    \includegraphics[width=\linewidth]{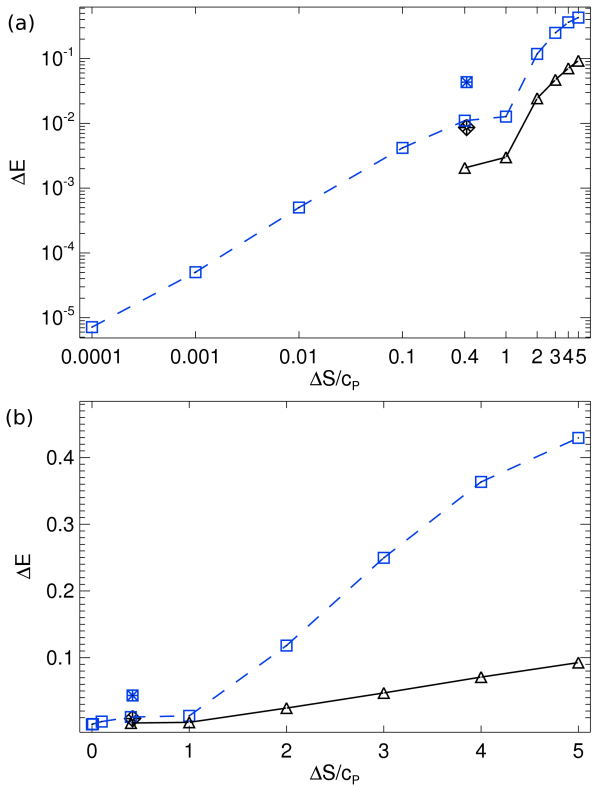}
\end{center}
  \caption{Violation of energy conservation in ASH-ANS
  simulations in various atmospheres.  
  $(a)$ Relative energy variation $\Delta E$ as given in
  equation~(\ref{eq:VE and VP}) for isothermal atmospheres with
  different non-dimensional energy drops $\Delta S/c_P$ in log-log
  plot. $(b)$ Same, in linear plot, emphasizing the behavior at large $\Delta S/c_P$.
  Shown in both are solutions with a single horizontal wave ($\ell=30$,
  blue squares) and solutions with many horizontal waves
  ($\ell=1$--$30$, black triangles).  Also shown are solutions in a
  solar radiative zone atmosphere stretching from 0.5--0.7$R_\odot$,
  with a single-wave solution (blue square with asterisk) and a many-wave solution
  (black diamond with asterisk).  All solutions are time-averaged over
  an interval of 2000$\tau_{BV}$, generally beginning about
  200$\tau_{BV}$ after the start of the simulation.
 \label{fig:many atmospheres and single waves}}
\end{figure}

The time-averaged relative energy variations in these ASH-ANS
simulations are shown in Figure~\ref{fig:many atmospheres and single waves}, 
which displays both many-wave solutions ($\ell=1$--$30$, triangles) and
single-wave solutions ($\ell=30$, squares).  The many-wave solutions
span from $\Delta S/c_P = 0.4$--$5$, while the single-wave solutions
span a wider range from $\Delta S/c_P=0.0001$--$5$.
As the stratification increases, energy non-conservation becomes increasingly
significant in the ASH-ANS equations, with $\Delta E$ approaching 10\%
in the many-wave simulation with $n_\rho=12.5$ and $\Delta S/c_P=5$ 
and $\Delta E \approx 43\%$ in the corresponding single-wave
solution.  Generally, we find that the relative energy variations are
about 5 times higher in the single-wave solutions than the
corresponding many-wave solutions, independent of stratification.  
In the corresponding ASH-LBR simulations
(not shown), energy is always well conserved with $\Delta E < 10^{-6}$.

Unsurprisingly, the relative energy variation is smaller in less
stratified atmospheres.  At very low levels of stratification 
($\Delta S/c_P \rightarrow 0$)
the ratio of pseudo-density and density is almost constant throughout
the domain (eq~\ref{eq:scaled pseudo-density}), and we
should expect the pseudo-energy conserving ASH-ANS equations to also
conserve energy fairly well.  Indeed, this is what we find.  As shown
for single-wave solutions in Figure~\ref{fig:many atmospheres and single waves}$(a)$, at low
values of $\Delta S/c_P$, the energy variation $\Delta E$ is also
small ($\Delta E \approx 5 \times 10^{-5}$ at $\Delta S/c_P=0.001$).
With increasing stratification, $\Delta E$ in the 
single-wave ASH-ANS solutions scales almost linearly
with $\Delta S/c_P$ up through $\Delta S/c_P=0.1$.  At $\Delta S/c_P=0.4$ 
$\Delta E\approx 1\%$ in the single-wave solution and $\Delta E\approx
0.2\%$ in the many-wave solution. 
There is a change in the scaling for both single-wave and many-wave
solutions at $\Delta S/c_P\approx 1$, apparent in both 
Figures~\ref{fig:many atmospheres and single waves}$(a,b)$; we do not
understand the origin of this behavior.
In all cases shown here, $\Delta PE \ll \Delta E$, 
and generally $\Delta PE \sim 10^{-5}$--$10^{-6}$.  This floor on
$\Delta PE$ likely reflects aspects of our data analysis technique and
we feel that our current approach is insufficient to reliably measure
energy variations in cases where $\Delta S/c_P \lesssim 10^{-4}$.

\subsection{Nonlinear interactions}
At much lower levels of diffusivity, or at larger initial amplitudes, 
the gravity waves may begin to interact nonlinearly. This is also
likely to occur when the gravity waves are driven by overshooting
convection from below \citep[e.g.,][]{Mihalas&Toomre_1981}.   
To confirm the linear nature of the waves we have studied here, we
define a Froude number $Fr$ as
\begin{equation}
  Fr = \frac{|\del \times \vec{u} |}{N},
\end{equation}
or the ratio of local vorticity to the Brunt-V\"ais\"al\"a frequency $N$.
This corresponds to the vorticity criteria for nonlinearity in 
\cite{Mihalas&Toomre_1981}.
We find here that $Fr$ attains a peak value of about $5\times10^{-5}$ in
the $n_\rho=7.5$ ASH-ANS simulation and of about $1 \times 10^{-5}$ in
the corresponding ASH-LBR simulation.  Thus the waves studied 
here are quite linear.  For linear waves, the Froude number gives the
characteristic amplitude of all fluctuations.  Owing to this, despite
the large stratifications studied here ($\Delta S/c_P \sim 1$), the
thermodynamic fluctuations remain quite small ($S_1/c_P \sim 10^{-5}$).

When nonlinear interactions become important, we might expect that
the non-conservation of energy may cause ASH-ANS simulations to
diverge even further from simulations which do conserve energy 
(e.g., the ASH-LBR equations). 
As discussed in Section~\ref{sec:ANS energy},
the conservation of pseudo-energy also vanishes when nonlinearity is
important.  Energy is conserved in the nonlinear LBR equations, but 
neither the pseudo-energy nor the energy is conserved
in the nonlinear ANS equations.  If the pseudo-energy is also not
conserved, it may be possible to \emph{inject} pseudo-energy into
otherwise closed systems; alternatively, the energy and pseudo-energy
may leak away without coupling to the reservoir of internal energy.
Either case leads to physical inconsistencies.  

Lastly, the transport by nonlinear processes in the ANS equations is
likely to be very different from that in equations that do conserve
energy, as the eigenfunctions of gravity waves in the ANS equations
are significantly higher in amplitude in the upper domain of
stably-stratified atmospheres
(Figures~\ref{fig:N7.5 eigenfunctions}--\ref{fig:ANS and LBR eigenfunctions}).
This will also have important implications for mode coupling, 
for the steepening and breaking of gravity waves, and for all other
problems where the shape of the eigenfunction itself is important.

\subsection{Solar atmospheres}
The results presented here so far have been for the special case of an
isothermal atmosphere.  The solar radiative zone is stably stratified,
but has fewer density scale heights than have been considered in most
of this section.  Across the entire solar radiative zone
\begin{equation}
 (\Delta S/c_p)_\odot = \int_{0.0R_\odot}^{0.7R_\odot} \frac{1}{c_P}\frac{\partial S_0}{\partial r}
 \approx 1.3, 
\label{eq:solar delta S}
\end{equation}
and $n_{\hat{\rho},\odot} \approx 7.9$, 
while $\Delta S/c_p \approx 0.42$ 
and $n_{\hat{\rho},\odot} \approx 2.3$ over the shell geometry
that we consider here ($0.5$--$0.7R_\odot$).

To constrain our results, we have repeated these gravity
wave rundown ASH-ANS and ASH-LBR simulations in the solar radiative
interior.  We take our model atmosphere from the CESAM code 
\citep{Brun_et_al_2002}.  We keep the same values of $\nu$ and
$\kappa$ and continue to neglect radiative diffusion acting on the
fluctuating flows.  A large scale radiative diffusion based on the
Rossland mean-opacity is included that acts on $\del T_0$ and 
there is a flux equal to the solar flux throughout the domain.  
We keep the same choice of $r_\mathrm{top}$ and $r_\mathrm{bot}$, thus 
$n_\rho \approx 1.8$ and $\Delta S/c_P\approx 0.417$.  

These solar simulations are shown in Figure~\ref{fig:many atmospheres and single waves}$(a)$
as asterisks.  In this model solar atmosphere, we find that in
many-wave solutions ($\ell=1$--$30$) with the ASH-ANS equations
$\Delta E \approx 0.9\%$ and $\Delta PE$ is tiny while in the ASH-LBR
equations $\Delta PE \approx 0.9\%$ and $\Delta E$ is tiny.  In
single-wave solutions ($\ell=30$), the relative energy variations 
in the ASH-ANS equations are much larger ($\Delta E \approx 4.5\%$).
Surprisingly, the solar atmosphere simulations show larger relative
energy variations than similarly stratified isothermal atmosphere
simulations, with $\Delta E$ being roughly five times larger
in this solar atmosphere than in the corresponding 
$\Delta S/c_P=0.4$  isothermal atmosphere.  
If we plotted these against $n_{\hat{\rho}}$, the
solar simulations would lie midway between the $\Delta S/c_P = 0.1$
and $0.4$ atmospheres and would still be clearly discrepant. 
We expect that the effects of energy non-conservation 
will become significantly larger as more entropy scale heights are
included in the domain.   This may be difficult to diagnose in
simulations that include a realistic solar stratification as the
radially varying Brunt-V\"ais\"al\"a frequency creates acoustic cavities that
may trap high frequency gravity waves, but we expect that the low
frequency waves which travel the entire radiative zone and experience
the full stratification will be affected.

\section{Recommendations for improving anelastic treatments of
    stellar interiors}
\label{sec:conclusions}

The results of Sections~\ref{sec:isothermal atmosphere}--\ref{sec:bounded geometries} 
provide a clear path to improving the treatment of dynamics within
stably-stratified atmospheres in anelastic systems of equations.
As clearly shown in Figure~\ref{fig:ANS and LBR E and PE}, the ANS
equations do not conserve energy and instead conserve a stratification
weighted pseudo-energy.
These equations thereby obtain incorrect frequencies and radial
eigenfunctions for gravity waves in both infinite and bounded
isothermal atmospheres.  The variation in the eigenfunctions is
substantially larger than the level of energy non-conservation.
These results hold in general for all
subadiabatically-stratified atmospheres.  

In contrast, the anelastic LBR equations do conserve energy and 
appear to need no additional modification to
capture dynamics in subadiabatically-stratified regions.  
This is fairly surprising, as those equations are generally derived in
nearly adiabatic atmospheres and the isothermal atmospheres we have
considered here take them far from their realm of validity.  At low
vertical and horizontal wave numbers, eigenfunctions in the LBR
equations differ from the full compressible equations, and results
from gravity waves in this regime should be treated with caution.  
These differences shrink as either wavenumber increases 
and the LBR equations may do a reasonable job of capturing the
dynamics of those shorter wavelength gravity waves 
(e.g., Figure~\ref{fig:N7.5 eigenfunctions}). 
Though we did not explore their dynamics in direct numerical simulation, we have
demonstrated that the RG equations do not conserve energy in general
atmospheres that have temperature gradients, and in those atmospheres
will also likely obtain incorrect radial eigenfunctions.

To correctly capture the dynamics in sub-adiabatically stratified
regions, it is vitally important that subsonic treatments of the
fluid equations conserve energy.  Systems of equations that conserve
energy are physically self-consistent, even if simulations done with
them have transport coefficients (e.g., $\nu$ and $\kappa$) that are
several orders of magnitude larger than the molecular values in
astrophysical systems.  Systems that do not conserve energy are not
physically consistent, and though the variations of energy may be
small for some problems, these variations point to deeper underlying
problems with those systems of equations.  In particular, the
eigenfunctions of the waves are significantly different in the
non-conservative systems (e.g., ANS) from the energy conserving
systems (e.g., FC and LBR), and this is very important for nonlinear
transport, mode coupling and wave steepening and breaking.
The clear path forward is to ensure that simulations employ anelastic
systems of equations that conserve energy; fortunately this can be
done with simple modifications to existing codes.

The route to energy conservation is to modify the momentum equation of
the non-conservative systems.  Fundamentally the conservation of
energy is more physical than the conservation of momentum: there are
many physical systems conserve energy instead of momentum, especially
those where a very fast restoring force acts to constrain the behavior
of the system.  Examples of this include inelastic scattering off of
rigid boundaries, where momentum changes sign but energy is conserved, 
and roller coasters on rigid tracks, where the track changes the
momentum of the careening roller coaster but not its total energy.
In anelastic systems, the fast sound waves provide the rapid restoring
force and the divergence constraint embodied in 
equation~(\ref{eq:anelastic continuity}) acts analogously to the rigid
tracks of the roller coaster, applying a continuous forcing to the
system.  Fundamentally, this forcing is energy-conserving in the LBR
equations but violates energy conservation in the ANS equations.

In simulation codes, there are two equivalent paths to convert the ANS
equations into an energy conserving form identical to the LBR equations.
The first path is by rewriting the equations to exactly match the
anelastic LBR equations.  This is done by solving for the reduced
pressure $\pomega$ instead of the fluctuating pressure $P_1$, and by
converting the buoyancy term to a ``codensity'' where entropy
fluctuations $S_1$ contribute to buoyancy but pressure fluctuations do
not.  Doing so causes the momentum equation to take the following form
\begin{equation}
  \frac{\partial \vec{u}}{\partial t} + \vec{u}\cdot\vec{\del}\vec{u} = 
  -\vec{\del}\pomega -\frac{S_1}{c_p}\vec{g} -\vec{\del}\cdot\vec{\scrD},
  \label{eq:fixed ASH momentum I}
\end{equation}
which is identical to the LBR momentum equation~(\ref{eq:ASH-LBR momentum}).

The second path to energy conservation is considerably simpler and
relies on introducing a correction term into the momentum
equation.  Generally, energy non-conservation occurs in anelastic
systems of equations when the reduced pressure $\pomega$ interacts
with the background stratification. The problematic term in the ANS
momentum equation~(\ref{eq:ASH momentum pomega}) is the 
$\pomega \vec{\del}(S_0/c_p)$ term.
We modify the momentum equation to read
\begin{equation}
  \frac{\partial \vec{u}}{\partial t} + \vec{u}\cdot\vec{\del}\vec{u} = 
  -\frac{1}{\rho_0}\vec{\del}P_1 + \left(\frac{P_1}{\gamma P_0} +F_\mathrm{BVZ}\right)\vec{g} 
  -\frac{S_1}{c_p}\vec{g} -\vec{\del}\cdot\vec{\scrD},
  \label{eq:fixed ASH momentum II}
\end{equation}
where the correction term is
\begin{equation}
  F_\mathrm{BVZ} = \frac{P_1}{g\rho_0} \del\left(S_0/c_P\right),
  \label{eq:F-BVZ correction term}
\end{equation}
and where equation~(\ref{eq:fixed ASH momentum II}) reduces to the LBR
momentum equation~(\ref{eq:ASH-LBR momentum}).
We remind the reader that we have taken $\vec{g} = -g \vec{\hat{r}}$, and this
sign is incorporated into equation~(\ref{eq:F-BVZ correction term}).
In a code like the ASH code, where the intermediate variable $\rho_1$
is carried around, this amounts to changing the equation of state for
density fluctuations to
\begin{equation}
\frac{\rho_1}{\rho_0} = \frac{P_1}{\gamma P_0} 
                                 + \frac{P_1}{g \rho_0}\del\left(\frac{S_0}{c_P}\right) 
                                 - \frac{S_1}{c_P} =
                                  \frac{P_1}{g \rho_0}\del\ln \rho_0
                                 - \frac{S_1}{c_P}.
\end{equation}
Equations~(\ref{eq:fixed ASH momentum I}) and (\ref{eq:fixed ASH momentum II})
are mathematically equivalent, but implementing this second path in a production code like ASH
requires only a few lines of code and is considerably simpler than
re-writing the equations in terms of $\pomega$.  We have implemented
both approaches in the ASH code, and the two paths to energy
conservation give identical results in these test simulations.

The RG equations appear to not conserve energy in any atmosphere that
has a temperature gradient.  Instead they conserve a pseudo-energy
weighted inversely by the background temperature $T_0$ in both
stably-stratified radiative zones and in nearly
adiabatically-stratified convection zones.  
This is an important distinction, as the differences between
pseudo-energy conservation and energy conservation can appear in the
RG equations even when $\Delta S/c_P$ is small.  Instead, what matters
is the number of temperature scale heights across the domain.  This
can be seen from the RG pseudo-density (\ref{eq:RG pseudo density})
and the associated pseudo-density scale height
\begin{equation}
 n_{\hat{\rho}} = n_\rho  - n_T ,
 \label{eq: N pseudo rho RG}
\end{equation}
where $n_T$ is the number of temperature scale heights.
(compare with eqn~\ref{eq: N pseudo rho arbitrary}).
In the RG equations, $n_T$ has a similar but opposite role as $\Delta S/c_P$ in
the ANS equations: increased $n_T$ leads to less pseudo-density stratification.
In the solar interior, $T_0$
is about $15 \times 10^6$K near the core, about $2\times 10^6$K at the
base of the convection zone ($0.7R_\odot$) and roughly 
$4 \times 10^5$K in the upper convection zone ($0.93R_\odot$).  This
leads to $n_T \approx 2$ and $n_{\hat{\rho}}\approx 4.7$ across the
solar radiative zone ($0.001$--$0.7R_\odot$), and to 
$n_T \approx 1.6$ and $n_{\hat{\rho}}\approx 0.9$ across the deep
convection zone ($0.7$--$0.93R_\odot$).
The energy conserving properties of the RG equations could be studied
in either a solar interior setting or in a polytropic atmosphere where
there is a linear temperature gradient \citep[e.g.,][]{Jones_et_al_2011}. 

As with the ANS equations, it is straight-forward to put the RG
equations into energy-conserving form.
The term that leads to energy non-conservation is the 
$\pomega \del \ln T_0$ term in equation~(\ref{eq:RG momentum pomega}),
which arises from a correction term,
\begin{equation}
\frac{1}{T_0}\frac{P_1}{g \rho_0}\frac{\partial T_0}{\partial r} g\vec{\hat{r}} = 
\pomega \vec{\del} \ln T_0,
\label{eq:RG correction term}
\end{equation}
intended to more correctly capture sub-adiabatic stratifications \citep{Rogers&Glatzmaier_2005_ApJ}.
If the RG momentum equation were re-written as
\begin{equation}
  \frac{\partial \vec{u}}{\partial t} + \vec{u}\cdot\vec{\del}\vec{u} = 
    - \vec{\del} \left(\frac{P_1}{\rho_0}\right) 
   +  \frac{T_1}{T_0} g \vec{\hat{r}},
    \label{eq:RG momentum corrected}
\end{equation}
then these equations would conserve energy, though the frequencies of
gravity waves in these equations remain a factor of $\sqrt{\gamma}$
higher than both the LBR frequencies and the low-frequency branch of
the full compressible Euler equations.  The source of this remaining
disagreement remains unclear.  It would be very interesting to see whether the
non-conservation of energy has any impacts on the nature of convection
in the RG equations, and on the coupling of convection to
stably-stratified regions above and below.

Here we have explored how gravity waves in a solar radiative interior
may be affected by anelastic treatments.  The Boussinesq equations,
which we have not considered here, are well known to conserve energy
both linearly and nonlinearly, and this indicates that the issue is
not the filtering of sound waves alone.  Rather, it is the treatment
of filtered, subsonic motions in a stratified atmosphere 
(in Boussinesq treatments the background density is constant).
Other treatments of subsonic motions, such as the 
pseudo-incompressible equations and Reduced Sound Speed Techniques
\citep[e.g.,][]{Rempel_2005, Rempel_2006, Hotta_et_al_2012} may
similarly not conserve energy, and we would suggest that
this be carefully tested.  Generally, the isothermal atmospheres
considered here or similar stably-stratified polytropic atmospheres
(not considered here) provide simple test cases.  The next paper in
this series will consider variations on the pseudo-incompressible
equations.

The subsonic dynamics of gravity waves may play an
important role within the radiative envelopes in more massive stars,
such as main-sequence A-, B- and O-type stars, where convective
overshoot drives gravity waves up into a rarifying envelope leading to
possible nonlinear wave breaking.  
Taking a CESAM model of a $2M_\odot$ A-type star, we estimate that the
entropy change across the radiative envelope is about
\begin{equation}
  (\Delta S/c_P)_{2M_\odot} = \int_{0.15R_*}^{0.97R_*}
  \frac{1}{c_P}\frac{\partial S_0}{\partial r} \approx 3.8,
\end{equation}
which is substantially larger than the drop across the solar radiative zone
(eq~\ref{eq:solar delta S}).  Over this range of radii, $n_\rho\approx
16$ and $n_{\hat{\rho}}\approx 20$.  Though we have focused on stellar
interiors, energy conservation within anelastic systems is an
important concern for dynamics in any stably-stratified atmosphere,
including planetary atmospheres, planetary interiors, and
astrophysical accretion disks \citep[e.g.,][]{Barranco&Marcus_2005}.
Energy conservation in anelastic equations may also play an important
role  when magnetic fields are included in questions of the dynamics of
magnetohydrodynamic instabilities including magnetic buoyancy
instabilities \citep{Berkoff_et_al_2010}.

Conservation of energy remains among the most sacrosanct and useful of principles in the physicist's      
toolbox.  The principle of energy conservation applies not only to ideal systems but furthermore to       
fundamentally dissipative or externally driven situations.  In the latter scenarios, a time-dependent     
statement of energy budget replaces the simpler notion of time-constancy of total energy in an            
isolated ideal system.  In all situations, we believe that one should not tolerate the existence of uncontrolled        
spurious kinetic sources.  For the particular problem of gravity and acoustic waves, this principle is    
more than philosophical.  When examining the gravity and acoustic waves, energy-conserving anelastic      
models reproduce fully compressible results with much greater fidelity than energy-violating anelastic    
models; and this fact produces implications for our understanding of stellar interiors.  Our particular       
work further shows that the existence of a conserved ``pseudo-energy''
does not rescue the energy non-conserving models.         
Rather, the existence of the pseudo-energy merely indicates why
these problems appear to have gone unnoticed in previous  simulations.  
However, including any nonlinear or dissipative effects leads to the impossibility of       
even proper pseudo-energy budgeting; in this case dissipation can even \emph{inject} positive pseudo-energy.     
There is in short no way around the issue.  Our best advice: properly account for energy whenever         
possible.

\acknowledgements
We thank Mark Miesch for his help in implementing the LBR equations in
the ASH code (by more difficult path one).   We thank Matthew Browning and Allan Sacha
Brun for supplying the CESAM A-type star model.  We thank Fausto Cattaneo for 
useful discussions about time-stepping errors.  We thank Daniel Lecoanet,
Mark Rast, Eliot Quataert and Gary Glatzmaier for reading this paper
and making useful suggestions.  We additionally thank Ann Almgren, Nic
Brummell, Chris Jones, Jon Dursi, Keith Julien, Tami Rogers, 
Steve Tobias, Juri Toomre, Toby Wood and everyone previously thanked for useful
discussions while exploring these anelastic equations.  We thank the
anonymous referee for their careful read of the paper.  Ellen Zweibel thanks the
Department of Astronomy at U.\ Chicago for their hospitality; a
portion of this work was completed there.  Benjamin Brown is supported in part by NSF
Astronomy and Astrophysics postdoctoral fellowship AST 09-02004.  
CMSO is supported by NSF grant PHY 08-21899.  The simulations were carried
out with NSF PACI support of NICS and TACC.

\appendix
\section{Conservative form of buoyancy term}
\label{sec:appendix conservative buoyancy}

When analyzing energy conservation properties in the anelastic equations (Section~\ref{sec:self-adjointness}),
the energy equation takes the general form
\begin{equation}
\pd{K}{t} + \div \left[\,\vec{u} (K + \rho_{0} \varpi )\,\right] + \rho_{0}\vec{u} \dot \vec{g} \, \frac{ S_{1}}{c_{p}} = RHS ,
\label{eq:LHS anelastic momentum}
\end{equation}
with kinetic energy density $K=\rho_0 u^2/2$ 
and where $RHS$ is the right hand side (e.g., eqns~\ref{eq:ASH energy} and \ref{eq:LBR energy-0}).  
In equation~(\ref{eq:LHS anelastic momentum}), the buoyancy work term
takes the form $\rho_{0} \vec{u} \dot \vec{g} S_{1}$.
We put this buoyancy work term into conservative form by using the
gravitational potential, $\vec{g} = - \grad \Phi$ and by recognizing
the relation
\begin{equation}
\rho_{0} \vec{u} \dot \vec{g} S_{1} = 
-\partial_{t} \left(\rho_{0}\Phi S_{1}\right) - \div \left(\rho_{0} \vec{u} \Phi S_{1}\right) - \Phi \,\rho_{0} \vec{u} \dot \grad S_{0}.
\label{eq:conservative buoyancy work I}
\end{equation}
Using equation~(\ref{eq:conservative buoyancy work I}), the left hand
side (eqn.~\ref{eq:LHS anelastic momentum}) can be put into
conservative form
\begin{eqnarray}
   \pd{E}{t} + \div \left[\,\vec{u} (E + \rho_{0} \hat{\varpi})\,\right] = RHS
\end{eqnarray}
where 
\begin{eqnarray}
&& E = \rho_{0}\left(\frac{|\vec{u}|^{2}}{2} - \Phi
  \frac{S_{1}}{c_{p}}\right),\label{eq:anelastic energy E I}\\
&& \hat{\pomega} = \pomega - \frac{1}{c_{p}}\int_{a}^{r} \Phi(r^\prime) \, \mathrm{d}S_{0}(r^\prime),
\label{eq:pomega hat}
\end{eqnarray}
with $a$ an arbitrary reference radius (here the radius of the lower
boundary) and where the entropy profile is monotonic. 

The energy defined in equation~(\ref{eq:anelastic energy E I}) is
slightly unusual in that the buoyancy contribution is not quadratic in
entropy perturbation $S_1$.  This can be put in a more familiar
quadratic form by first noting that for arbitrary (possibly
nonlinear) motions 
\begin{multline}
\rho_{0}\, \vec{u} \dot \vec{g} \, \frac{S_{1}}{c_{p}} = -\rho_{0}\frac{d\Phi}{d S_{0}}\, \frac{S_{1}}{c_{p}} \vec{u}  \dot \grad S_{0} = \frac{\rho_{0}}{2 c_{p}}\frac{d\Phi}{d S_{0}} \frac{D S_{1}^{2}}{Dt} = \\
\partial_{t} A + \div \left( \vec{u} \, A \right) - A \vec{u} \dot \grad \ln \left( \frac{d\Phi}{d S_{0}} \right), 
\label{eq:quadratic potential energy} 
\end{multline}
where the available potential energy $A$ is given by
\begin{eqnarray}
A = \frac{1}{2}\frac{\rho_{0}}{c_{p}}\frac{d\Phi}{d S_{0}} \,S_{1}^{2},
\label{eq:A}
\end{eqnarray}
and 
\begin{eqnarray}
\frac{d\Phi}{d S_{0}} = \frac{g(r)^{2}}{c_{p} N(r)^{2}}.
\label{eq:Dphi/DS0}
\end{eqnarray}
Equations~(\ref{eq:quadratic potential energy}--\ref{eq:Dphi/DS0}) let
us rewrite equation~(\ref{eq:LHS anelastic momentum}) as
\begin{eqnarray}
   \pd{\tilde{E}}{t} + \div \left[\,\vec{u} (\tilde{E} + \rho_{0}
     \hat{\varpi})\,\right]  - A\vec{u}\dot\vec{\del}\ln \left( \frac{d\Phi}{d S_{0}} \right)= RHS ,\nonumber\\
\label{eq:anelastic energy E I full}
\end{eqnarray}
where the alternative total energy $\tilde{E}$ is
\begin{eqnarray}
&& \tilde{E} \equiv \frac{1}{2}\rho_{0} |\vec{u}|^{2} 
+\frac{\rho_0}{2 c_{p}}\frac{d\Phi}{d S_{0}} \,S_{1}^{2} ,
\label{eq:anelastic energy E II}
\end{eqnarray}
and where $\hat{\pomega}$ is given by equation~(\ref{eq:pomega hat}).

The left-hand side of equation~(\ref{eq:anelastic energy E I full})
can be put into conservative form if the condition
\begin{equation}
A\vec{u}\dot\vec{\del}\ln \left( \frac{d\Phi}{d S_{0}} \right)=0
\label{eq:A condition}
\end{equation}
is satisfied.  This happens under under two different conditions:
(\textit{i}) if $d\Phi/dS_{0}$ is constant, or
(\textit{ii}) we only consider linear perturbations.  
If condition (\textit{i}) is satisfied (e.g., in isothermally- or
adiabatically-stratified atmospheres) then \eq{eq:A condition} 
holds for nonlinear motions as well and systems of equations with
$RHS=0$ will conserve a quadratic potential energy for nonlinear as
well as linear motions.

\section{Eigenfunctions for a bounded atmosphere}
\label{appendix:eigenfunctions}
Our analytic approach is similar to that in an infinite isothermal
atmosphere, except now the wavelike perturbations are expanded in
spherical harmonics and the radial eigenfunctions must be solved for.
We take the spherical shell geometry of Section~\ref{sec:bounded
  geometries} and take impenetrable boundary conditions at the upper
and lower boundary
\begin{equation}
\xi_r = 0~\text{at}~r=a, b
\end{equation}
where $a=r_\mathrm{bot}$ and $b=r_\mathrm{top}$ (and see Table~\ref{table:atmospheres}).
Analytic eigenfunctions can be found if we consider a simplified
atmosphere with constant gravity $\vec{g} = -g \vec{\hat{r}}$ and
constant Brunt-V\"ais\"al\"a frequency $N$, and we do so here as well as
in the main body of the text.

\subsection{LBR eigenfunctions}
We begin with the LBR equations. 
In this system, in a spherical shell geometry, equation~(\ref{eq:pomega horizontal divergence}) for
reduced pressure $\pomega$ becomes 
\begin{equation}
  \frac{\ell(\ell +1)}{r^2} \pomega = 
  -\omega^2 \left[ \xi_r H^{-1} -
    \frac{1}{r^2}\frac{\partial}{\partial r} (r^2 \xi_r)\right],
\end{equation}
where we have used the anelastic continuity
equation~(\ref{eq:anelastic continuity}).
Defining
\begin{equation}
  \phi(r) \equiv \xi_r(r) r e^{\left(-r/2H\right)}
\end{equation}
and
\begin{equation}
  \lambda \equiv -\ell(\ell + 1) \left(1 - N^2/\omega^2\right) ,
  \label{eq:lambda definition}
\end{equation}
the momentum equation~(\ref{eq:LBR vertical momentum omega}) becomes
\begin{equation}
  -\frac{\partial}{\partial r} \left( r^2 \frac{\partial}{\partial r} \phi(r) \right)
  + \frac{r^2}{4H^2}\phi(r) = \lambda \phi(r).
\end{equation}
For different $\lambda$ the different $\phi(r)$ are orthogonal and
this can be used to determine
\begin{equation}
  \lambda = \frac{1}{4} + k^2,
\end{equation}
where the vertical wavenumber $k$ is normalized by the pressure 
and density scale height $H$.
With equation~(\ref{eq:lambda definition}) we obtain the dispersion
relationship
\begin{equation}
  \omega^2 = \frac{\ell(\ell + 1)}{\ell(\ell + 1) + k^2 + \case{1}{4}}N^2 .
\end{equation}

The vertical wavenumber $k_n$ can be approximated as
\begin{equation}
  k_n^2 = \frac{n^2 \pi^2}{\ln\left(\case{b}{a}\right)^2}\left(1 +
    \frac{b^2 -a^2}{8 H^2} \frac{\ln\left(\case{b}{a}\right)}{n^2\pi^2
      + \ln\left(\case{b}{a}\right)^2}\right) + \scrO(H^{-4})
  \label{eq:approximate wavenumber}
\end{equation}
for a spherical shell with lower boundary at $r=a$ and upper 
boundary at $r=b$.  
Exact solutions can be found numerically by solving the problem in
terms of Bessel functions, with
\begin{multline}
  \xi_{r, \mathrm{LBR}}(r) = \Big[K_{ik}\left(\frac{a}{2H}\right) I_{ik}\left(\frac{r}{2H}\right)  \\
                     -  I_{ik}\left(\frac{a}{2H}\right) K_{ik}\left(\frac{r}{2H}\right)\Big] 
                   r^{-3/2} \exp{\left(\frac{r}{2H}\right)} ,
\end{multline}
where $I_{ik}$ and $K_{ik}$ are modified Bessel functions of the first
and second kind respectively with imaginary index $ik_n$.  The
impenetrable boundary conditions at $r=b$ requires that $\xi(r)=0$ and
thus
\begin{equation}
   K_{ik}\left(\frac{a}{2H}\right) I_{ik}\left(\frac{b}{2H}\right) 
-  I_{ik}\left(\frac{a}{2H}\right) K_{ik}\left(\frac{b}{2H}\right) =0,
\end{equation}
which can be solved by Newton's method and using
equation~(\ref{eq:approximate wavenumber}) as an initial guess,
yielding $k_n$.  

\subsection{ANS eigenfunctions}
At this point the eigenfunctions for the linear ANS equations can be
found by a simple transformation
\begin{equation}
  \xi \rightarrow  \xi \exp{(-S_0/c_P)},
\end{equation}
\begin{equation}
  H  \rightarrow  \gamma H,
\end{equation}
which leads to
\begin{equation}
   \pomega \rightarrow \pomega \exp{(-S_0/c_P)},
\end{equation}
and transforms the linearized LBR wave equations~(\ref{eq:LBR vertical momentum omega}) into the linearized
ANS wave equations~(\ref{eq:ASH vertical momentum omega}).  This transformation leads to eigenfunctions of
\begin{multline}
  \xi_{r,\mathrm{ANS}}(r) = \Big[K_{ik}\left(\frac{a}{2\gamma H}\right) I_{ik}\left(\frac{r}{2\gamma H}\right)\\
                     -  I_{ik}\left(\frac{a}{2\gamma H}\right) K_{ik}\left(\frac{r}{2\gamma H}\right)\Big] 
                   r^{-3/2} \exp{\left((2\gamma-1)\frac{r}{2\gamma H}\right)}.
\end{multline}
The vertical wavenumbers are found as before by solving
\begin{equation}
K_{ik}\left(\frac{a}{2\gamma H}\right) I_{ik}\left(\frac{b}{2\gamma H}\right)
                     -  I_{ik}\left(\frac{a}{2\gamma H}\right) K_{ik}\left(\frac{b}{2\gamma H}\right)=0.
\end{equation}

\subsection{RG eigenfunctions}
The linear RG are already in almost the same form as the LBR
equations, except that
\begin{equation}
  \lambda_{RG} \equiv -\ell(\ell + 1) \left(1 - \gamma N^2/\omega^2\right) 
  = \case{1}{4} + k^2.
  \label{eq:lambda definition RG}
\end{equation}
Thus the solutions for radial wavenumber $k$ and the radial
eigenfunctions are the same as in the LBR equations, but the
frequencies $\omega$ are a factor of $\sqrt{\gamma}$
higher than the Brunt-V\"ais\"al\"a frequency.

\subsection{Normalization of eigenfunctions}
As defined so far, the amplitude of the eigenfunctions $\xi$ is a free parameter.
In all sets of equations, we normalize the eigenfunctions by an
amplitude $A$, with 
\begin{equation}
A^2 = \frac{\int^b_a{\xi(n, r)^2 \exp[-\epsilon r/H] r^2 dr}}  {\int^b_a{\exp[-\epsilon r/H] r^2 dr}} ,
\end{equation}
where $\epsilon$ represents the imaginary part of the vertical
wavenumber $K$ and is
\begin{equation} 
\epsilon=
\begin{cases} 
  2 - (1/\gamma) & \text{ANS equations}\\ 
  1 & \text{all others}.
\end{cases} \end{equation}
This choice of normalization gives the correct amplitude for motions
in the different systems of equations when subject to the same initial
conditions (entropy perturbations of fixed initial amplitude); this is
how we conduct the 3-D numerical simulations 
and thus the analytic eigenfunctions shown
in Section~\ref{sec:isothermal analytic bounded} show the same
amplitude ordering as the numerical simulations of 
Section~\ref{sec:ASH sims}.



\end{document}